\documentclass[12pt,a4paper,onecolumn]{article}
\ifdefined\pdfminorversion\pdfminorversion=7\relax\fi
\newif\ifdraftnotes
\draftnotesfalse
\newif\ifreviewlines
\reviewlinesfalse
\usepackage{iftex}
\ifPDFTeX
  \usepackage[T1]{fontenc}
  \usepackage[utf8]{inputenc}
\fi
\usepackage{arxiv}
\usepackage{amsmath,amssymb,mathtools}
\usepackage{microtype}
\usepackage{lineno}

\usepackage{graphicx}
\graphicspath{{figures/}}
\usepackage{xcolor}
\definecolor{gpanote}{HTML}{31596B}
\definecolor{gpaframe}{gray}{0.65}
\definecolor{gpaframetext}{gray}{0.35}
\usepackage{caption}
\usepackage{subcaption}
\usepackage{booktabs,array,tabularx,multirow}
\newcolumntype{Y}{>{\raggedright\arraybackslash}X}
\usepackage{siunitx}
\usepackage[version=4]{mhchem}
\usepackage{algorithm}
\usepackage{algpseudocode}
\algrenewcommand\algorithmicindent{1em}
\usepackage{placeins,flafter}
\usepackage[numbers,sort&compress]{natbib}
\usepackage{xurl}
\usepackage{xr-hyper}
\usepackage[hidelinks]{hyperref}
\IfFileExists{supplement.aux}{\externaldocument[][nocite]{supplement}}{
\externaldocument[][nocite]{supplement_xr}}
\usepackage[nameinlink,noabbrev,capitalise]{cleveref}
\crefname{figure}{Figure}{Figures}
\crefname{table}{Table}{Tables}
\crefname{equation}{Equation}{Equations}
\crefname{section}{Section}{Sections}
\crefname{algorithm}{Algorithm}{Algorithms}

\DeclareMathOperator*{\argmin}{arg\,min}

\newcommand{\LTP}{\ce{LiTi2(PO4)3}}
\newcommand{\LPS}{\ce{Li4P2S6}}
\newcommand{\LGPS}{\ce{Li10GeP2S12}}

\newcommand{\papertitle}{GeoSaddle: Geometry-Guided Saddle Networks for Periodic Ion Migration}
\newcommand{\papershorttitle}{GeoSaddle}
\newcommand{\paperpdftitle}{GeoSaddle: Geometry-Guided Saddle Networks for Periodic Ion Migration}
\newcommand{\paperkeywords}{ion migration; transition states; geometric path analysis;
machine learning interatomic potentials; solid electrolytes; periodic migration networks}
\hypersetup{pdftitle={\paperpdftitle},pdfauthor={Chengrui Xu and Shyue Ping Ong},
  pdfsubject={Preprint},
  pdfkeywords={ion migration, transition states, geometry, periodic networks}}

\renewcommand{\shorttitle}{\papershorttitle}
\renewcommand{\headeright}{Preprint}
\renewcommand{\undertitle}{Preprint}

\makeatletter
\providecommand{\theHALG@line}{}
\renewcommand{\theHALG@line}{\thealgorithm.\arabic{ALG@line}}
\renewcommand{\maketitle}{%
  \begingroup
  \renewcommand{\thefootnote}{\fnsymbol{footnote}}%
  \let\Hy@saved@footnotemark\@footnotemark
  \let\Hy@saved@footnotetext\@footnotetext
  \let\@footnotemark\H@@footnotemark
  \let\@footnotetext\H@@footnotetext
  \@maketitle%
  \thispagestyle{plain}%
  \@thanks
  \endgroup
  \setcounter{footnote}{0}%
}
\renewcommand{\@maketitle}{%
  \begingroup
  \centering
  {\LARGE\bfseries\@title\par}
  \vspace{6pt}
  {\small\textsc{\undertitle}\par}
  \vspace{10pt}
  \def\And{\end{tabular}\hspace{1.8em}\begin{tabular}[t]{c}\bfseries}
  \def\AND{\end{tabular}\par\vspace{6pt}\begin{tabular}[t]{c}\bfseries}
  {\small\begin{tabular}[t]{c}\bfseries\@author\end{tabular}\par}
  \vspace{14pt}
  \endgroup
}
\renewenvironment{abstract}{%
  \par\noindent{\large\bfseries Abstract}\par\vspace{5pt}
  \begingroup\normalsize\noindent\ignorespaces
}{\par\endgroup\vspace{5pt}}
\makeatother

\usepackage{supertabular}
\usepackage{needspace}

\title{\papertitle}
\author{%
  Chengrui Xu\\
  \normalfont Tanwei College\\
  \normalfont Tsinghua University\\
  \normalfont xucr23@mails.tsinghua.edu.cn
  \And
  Shyue Ping Ong\thanks{Corresponding author: shyue@nus.edu.sg}\\
  \normalfont Department of Materials Science and Engineering\\
  \normalfont National University of Singapore\\
  \normalfont shyue@nus.edu.sg
}
\date{}

\begin{document}
\maketitle

\begin{abstract}
Ion migration mechanisms govern ionic conductivity in materials such as solid electrolytes and electrodes for lithium-ion batteries. However, nudged elastic band (NEB) calculations require the endpoints and an initial path to be specified in advance, which precludes the discovery of multi-step and concerted mechanisms whose intermediate states are unknown. Here, we present GeoSaddle, which uses periodic geometric accessibility and channel directions to initialize local saddle searches on a potential energy surface (PES). Verified saddle--minimum connections supply actual product configurations for subsequent searches, and matching atomic occupancies across periodic images assembles these events into migration networks beyond a single prescribed endpoint pair. We evaluate GeoSaddle on 100 paths from the LiTraj database using a foundation potential (FP). On ten long, curved events with known near-reference saddles, GeoSaddle required 12.7\% of the total model evaluations used by an ellipsoidal PES-grid control. GeoSaddle recovered 92 of 100 LiTraj paths, with a 38.9\% lower mean path root-mean-square (RMS) deviation than bond-valence site energy (BVSE) guesses on the recovered set (0.120 versus 0.196~\AA), and reduced the number of subsequent FP NEB steps by 28\% relative to BVSE initialization. Beyond single endpoint pairs, GeoSaddle reconstructed published vacancy, split-interstitial and knock-out mechanisms into periodic relay sequences in \LTP{}, \ce{Li_{1+x}Al_{x}Ti_{2-x}(PO4)3} (LATP) and \LPS{}. In LATP, quenched \textit{ab initio} molecular dynamics configurations matched GeoSaddle network states, including branches outside the selected chain. In \LGPS{}, three-Li and four-Li periodic relay cycles identified by GeoSaddle preserved their transition connectivity under full-ion density functional theory (DFT) refinement, with peaks of 0.346 and 0.116~eV above their common initial state. In pyrochlore \ce{Li_{2-x}La_{(1+x)/3}Nb2O6F}, sublattice exploration resolved a six-step, three-Li relay with a 0.218~eV full-ion FP NEB peak and isolated the barrier reduction from local fluoride relaxation. GeoSaddle requires only energies and forces, and its open-source implementation provides a general route from a host structure to verified, periodic migration networks.
\end{abstract}

\ifreviewlines\linenumbers\fi

\section{Introduction}
\label{sec:introduction}

Ion migration pathways and their associated barriers are central to understanding ionic transport~\cite{he2017concerted}. A common approach to compute ion migration barriers is the nudged elastic band (NEB), wherein a set of configurations between specified initial and final states is optimized using perpendicular physical forces and tangential spring forces~\cite{henkelman2000ci} to identify the minimum energy path (MEP). The speed of convergence and, in some instances, the final result are highly sensitive to the initial guess of the MEP, including the identification of the initial and final states~\cite{smidstrup2014idpp,rong2016pathfinder}.

The most common approach to create the initial guess of the MEP is simple linear interpolation between the initial and final states. However, several approaches have been developed to generate improved, physically informed MEP guesses using approximations to the potential energy surface (PES). For example, the bond-valence site energy (BVSE) method guides path construction using an empirical bond-valence energy field~\cite{adams2011bvse}. The image-dependent pair potential (IDPP)~\cite{smidstrup2014idpp} and crystal structure analysis by Voronoi decomposition (CAVD)~\cite{he2020cavd} methods construct the MEP guess using geometric arguments such as interatomic distances. \citet{rong2016pathfinder} also proposed an approach that uses density functional theory (DFT) charge densities to optimize the MEP guess. More recently, machine learning interatomic potentials (MLIPs) have been used to optimize MEP guesses using a learned PES~\cite{dembitskiy2025litraj}. Despite this progress, all of these approaches still require the initial and final states to be specified in advance. This requirement is most limiting for many materials with high ionic conductivity, such as the lithium superionic conductors \LGPS{} (LGPS) and \ce{Li_{1+x}Al_{x}Ti_{2-x}(PO4)3} (LATP)~\cite{he2017concerted}, in which the lowest-barrier mechanisms often involve the concerted motion of multiple ions through intermediate states that are not known a priori.

Because macroscopic transport requires percolating sequences of such events, a second challenge is to assemble individual transitions into networks of minima and transition states. While various algorithms and workflows have been proposed to identify transition state networks~\cite{xu2026moto,jung2026nebscape,barkema1996art,malek2000art,elmellouhi2008kart,beland2011kart}, they share a similar underlying mathematical framework: constructing saddle--minimum (Morse--Smale) energy networks to map localized configuration landscapes. However, each is tailored to other atomistic problems, such as surface reactions or event catalogues for off-lattice kinetic Monte Carlo, rather than ion migration in crystals. Applying these methods to ion migration requires two further capabilities: seeding saddle searches efficiently within the migration channels of the host crystal, and composing local transitions across periodic images to identify propagating and concerted mechanisms.

Here, we introduce GeoSaddle, which addresses both challenges: finding migration paths with transition states efficiently and assembling them into periodic transport networks. GeoSaddle couples periodic geometric accessibility to local saddle searches on a PES. A purely geometric analysis of the host crystal identifies candidate positions and migration directions within accessible channels, which concentrates the subsequent saddle searches on configurations relevant to migration. GeoSaddle requires only energies and forces; here we use a foundation potential (FP), a universal MLIP trained across the periodic table, to evaluate the PES. Building on established saddle--minimum network methods, GeoSaddle continues searches from the actual product configurations of accepted events and composes atom-occupancy mappings and periodic image shifts into migration networks. This combination supports both endpoint-defined pathways and mechanism exploration as local occupancies change. Unlike existing transition-state network methods~\cite{xu2026moto,jung2026nebscape}, GeoSaddle seeds its searches from the channel geometry of the host and certifies net periodic transport through cycles of the state graph. We demonstrate the utility of GeoSaddle through a large-scale analysis of 100 paths from the LiTraj database (LiTraj-100)~\cite{dembitskiy2025litraj}, transfer to Na hosts and tests of NEB initialization. We then test periodic mechanism reconstruction in the model Li conductors \LTP{} (LTP), LATP and \LPS{} (LPS), followed by collective pathway exploration in LGPS and the pyrochlore oxyfluoride \ce{Li_{2-x}La_{(1+x)/3}Nb2O6F} (LLNOF). We used DFT climbing-image NEB for the \ce{Li5BO4} benchmark and the LGPS relays, and \textit{ab initio} molecular dynamics for LATP. The LLNOF relay was refined with MatPES CI-NEB.

\begin{figure*}[!t]
  \centering
  \includegraphics[width=\textwidth]{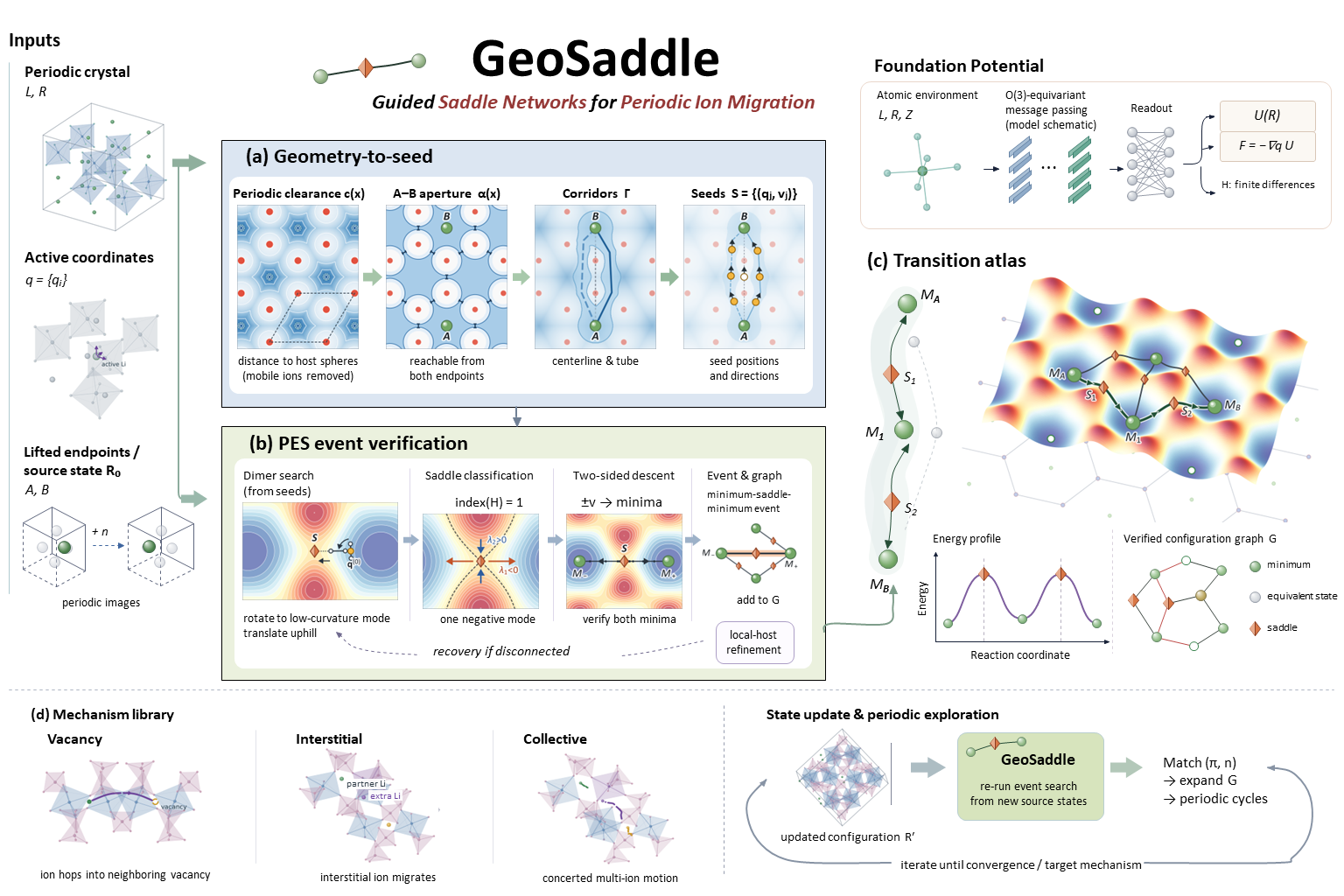}
  \caption{\textbf{Overview of the GeoSaddle framework.} (a) Periodic crystal clearance and endpoint-conditioned bottleneck aperture identify accessible channels and directional seeds without PES evaluations. (b) Local dimer saddle searches, active-space Hessian classification and two-sided relaxation establish verified minimum--saddle--minimum events. (c) Equivalent stationary points are merged into a local transition atlas from which minimum-barrier pathways are extracted. (d) State-to-state continuation from product configurations and periodic occupancy matching assemble vacancy, interstitial and collective rearrangements into propagating migration networks.}
  \label{fig:gpa-architecture}
\end{figure*}

\section{Results}
\label{sec:results}

GeoSaddle is constructed in four stages that connect host-crystal geometry to stationary-point networks on any PES (Fig.~\ref{fig:gpa-architecture}). First, distances to the nearest framework atoms define periodic clearance channels and bottleneck apertures in the host crystal that supply a trial position and initial mode vector as direction, termed a ``seed'', for each saddle search \textit{without} PES evaluations (Fig.~\ref{fig:gpa-architecture}a). Second, each seed initializes a single-ended dimer search~\cite{henkelman1999dimer} using a pre-trained FP to locate an index-one saddle point, and downhill relaxation from both sides of the saddle connects it to the adjacent minima through continuous pathway structures (Fig.~\ref{fig:gpa-architecture}b). In this work, the \texttt{TensorNet-PES-MatPES-PBE-2025.2} FP~\cite{kaplan2025matpes,simeon2023tensornet,ko2025matgl} is used for PES evaluations. Third, periodically equivalent stationary points are merged into a local transition atlas to extract the MEP and intermediate basins between specified endpoints (Fig.~\ref{fig:gpa-architecture}c). Finally, searches continue from the relaxed product configurations of accepted events to build extended periodic migration networks across unit cells (Fig.~\ref{fig:gpa-architecture}d).

To balance screening speed against framework accommodation, the primary searches move only the migrating ions and keep the host framework fixed. We refer to the coordinates that are allowed to move as the active space. Framework flexibility is then incorporated in larger active spaces through local-host saddle refinement, conditional host relaxation at fixed ion positions, coupled ion--framework verification and full-ion DFT relaxation. In the following sections, we evaluate each step of the GeoSaddle workflow: (i) geometry-guided saddle seeding, (ii) recovery of endpoint-defined pathways, (iii) NEB initialization and (iv) the reconstruction of periodic defect mechanisms. Finally, we extend GeoSaddle to the exploration of collective multi-ion pathways.

\subsection{Geometry-guided saddle search}
\label{sec:results-geometry}

\begin{figure*}[!tbp]
\centering
\includegraphics[width=.94\textwidth]{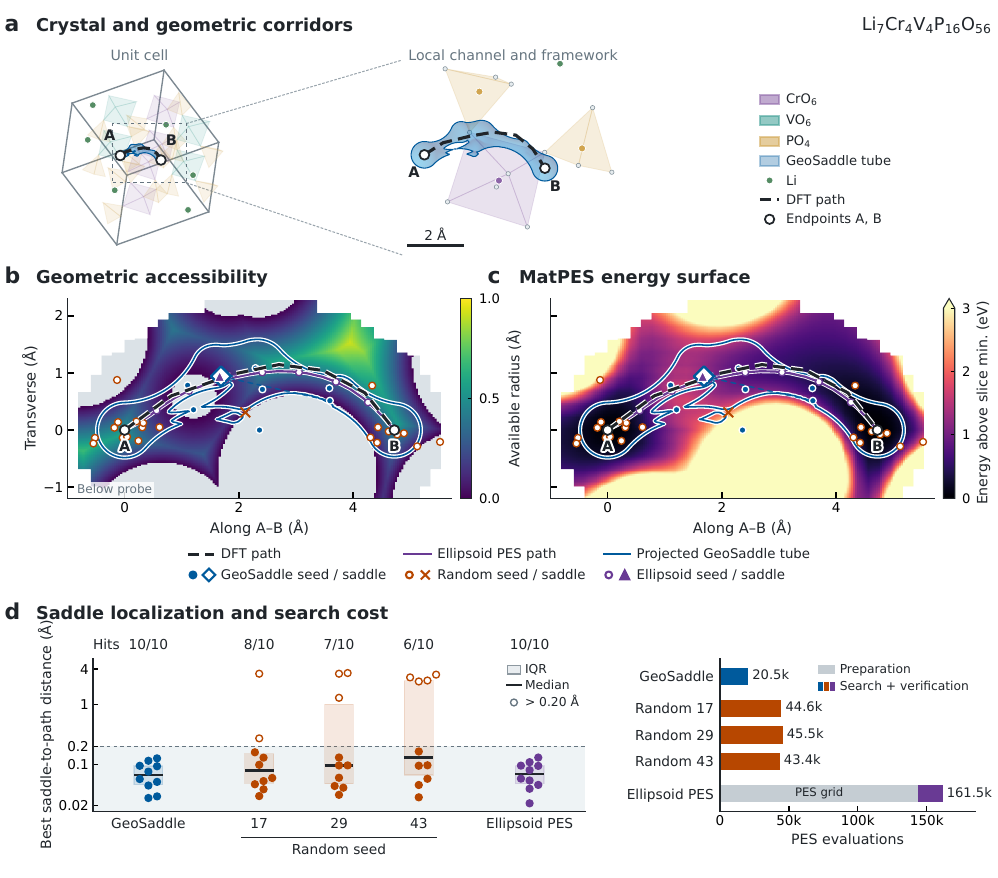}
\caption{\textbf{Geometric guidance for local saddle searches.}
(a) Unit-cell and local channel views of \ce{Li7Cr4V4P16O56} with the GeoSaddle tube and reference DFT path.
(b,c) Planar maps of available radius $r_{\mathrm{avail}}(\mathbf{x})=\max[c(\mathbf{x})-r_{\mathrm{p}},0]$ for periodic host clearance $c(\mathbf{x})$ and probe radius $r_{\mathrm{p}}=0.4$~\AA\ (b, with $c(\mathbf{x})\le r_{\mathrm{p}}$ in gray) and fixed-host MatPES energy above the slice minimum (c). Both maps are overlaid with projected paths, tubes and seed-to-saddle displacements (dashed arrows).
Panels (b,c) show only points and complete seed--saddle links within the displayed window.
(d) Nearest strict-saddle distance to the continuous DFT path across ten long, curved paths with near-path hit counts out of 10 (left), and cumulative PES evaluations split into preparation and search plus verification from the complete evaluation records (right).}
\label{fig:geometry-guidance}
\end{figure*}

Single-ended dimer searches~\cite{henkelman1999dimer} climb to an index-one saddle from a seed defined by a trial position and an initial mode vector. Near a crystal minimum, unguided seeds often follow local vibrational or non-diffusive modes to unrelated saddles outside the migration channel. To test whether channel geometry concentrates the searches on transition states relevant to migration, we compared three seeding approaches on ten long, curved migration events from LiTraj-100, each in a distinct host structure (Fig.~\ref{fig:geometry-guidance}). The three approaches are (i) GeoSaddle channel seeds, which use positions and local tangents along geometric channels, (ii) local random displacements around the relaxed endpoints, using each displacement direction as the initial mode, and (iii) seeds sampled along paths on ellipsoidal MatPES energy grids, which use local tangents along these paths as seeds. For the third control, the two prescribed jump endpoints define the foci of a prolate ellipsoid with eccentricity $e=0.70$ and a Cartesian grid spacing of at most 0.20~\AA. The ten migration events span endpoint separations of 4.52--5.91~\AA\ and DFT path lengths of 5.22--6.99~\AA, with path elongations of 1.13--1.24 and transverse chord deviations of 0.81--1.50~\AA. All three methods received the same attempt budget for each case, which amounts to 97 dimer attempts in total over the ten cases (Sec.~\ref{sec:evaluation}).

To illustrate, a representative \ce{Li7Cr4V4P16O56} structure and its positive clearance channels are shown in Fig.~\ref{fig:geometry-guidance}a--c. The positive-clearance channel tracks the curved MatPES valley around the host polyhedra and places directional seeds that converge directly to the migration saddle. Across all ten events, GeoSaddle converged to strict index-one saddles in 85 of the 97 dimer attempts, of which 72 lay within 0.20~\AA\ of the continuous DFT path and covered all ten cases with median and maximum nearest-saddle distances of 0.066 and 0.125~\AA\ (Fig.~\ref{fig:geometry-guidance}d and Supplementary Table~\ref{tab:geometry-guidance}). Across three independent repetitions, local random seeding still converged to strict saddles in 64--71 of the 97 attempts, yet only 17--19 attempts lay within 0.20~\AA\ of the DFT path and covered six to eight of the ten cases per repetition because most searches diverted to off-channel saddles. The ellipsoidal PES-grid control also recovered all ten cases with 76 near-path hits and a median distance of 0.068~\AA, but required 161,537 PES evaluations in total, of which 143,848 were spent on the energy grid and 17,689 on dimer search and verification. Channel seeding in GeoSaddle required zero PES preparation calls and 20,487 PES evaluations in total, 12.7\% of the grid-control cost and less than half of the 43,415--45,535 evaluations used by random seeding.

\subsection{Recovery of saddle-connected pathways}
\label{sec:results-blind}


After identification of the saddles, the next step is two-sided downhill relaxation and the merging of equivalent basins to link local transitions into a complete pathway between the endpoints $A$ and $B$. We evaluated this assembly on LiTraj-100~\cite{dembitskiy2025litraj} in a blind setting, without access to the interior images of the reference paths. LiTraj-100 comprises 64 oxide, 16 sulfide, 12 halide and 8 nitride single-Li DFT paths in 93 crystal structures. The endpoint separations range from 0.53 to 5.91~\AA\ with a median of 3.28~\AA, and the path lengths range from 0.57 to 6.99~\AA\ with elongations of up to 1.27. The reference forward barriers are as high as 1.00~eV, with a median of 0.38~eV and a 90th percentile of 0.85~eV (Supplementary Fig.~\ref{fig:si-initialization} and Supplementary Table~\ref{tab:si-cohorts}).

With the host framework fixed during the primary search, GeoSaddle recovered a chain of verified index-one saddles connecting $A$ and $B$ for 92 of the 100 paths (Table~\ref{tab:blind-main} and Supplementary Table~\ref{tab:si-blind-outcomes}). It connected two more paths through candidate saddles with near-zero secondary curvature. Of the six paths that remained unconnected, two had endpoints that relaxed into the same energy basin of the fixed host. For the 92 recovered paths, we compared the GeoSaddle chains and the BVSE paths against the DFT reference. The mean whole-path root-mean-square (RMS) deviation from DFT was 0.120~\AA\ for GeoSaddle compared with 0.196~\AA\ for BVSE, a reduction of 38.9\%. More importantly, GeoSaddle reduced the 90th-percentile path error by 56.8\%, from 0.433~\AA\ for BVSE to 0.187~\AA. The GeoSaddle chains also covered 88.56\% of the reference DFT pathways within 0.20~\AA. The selected saddles also lay close to the DFT paths, with a mean saddle-to-path distance of 0.120~\AA, compared with 0.205~\AA\ for the BVSE saddle proxy. Local-host dimer refinement, in which the framework atoms within 4.0~\AA\ of the saddle were also relaxed, further reduced the mean distance to 0.098~\AA\ and the median from 0.093 to 0.064~\AA.

\begin{table}[!tbp]
\centering
\small
\setlength{\tabcolsep}{3pt}
\renewcommand{\arraystretch}{1.08}
\caption{\textbf{LiTraj-100 geometry and network recovery.} GeoSaddle recovered 92 of the 100 pathways. The mean, median and 90\%-tile numbers are reported for these 92 pathways for both BVSE and GeoSaddle.}
\label{tab:blind-main}
\begin{tabularx}{\columnwidth}{@{}>{\raggedright\arraybackslash}Xrrr@{}}
\toprule
Output & Mean & Median & 90\%-tile \\
\midrule\multicolumn{4}{@{}l}{\textit{Whole-path RMS (\AA)}} \\
BVSE & 0.1962 & 0.1440 & 0.4325 \\
GeoSaddle chain & 0.1198 & 0.1111 & 0.1869 \\
\midrule\multicolumn{4}{@{}l}{\textit{Saddle-to-DFT-polyline distance (\AA)}} \\
BVSE proxy & 0.2049 & 0.1527 & 0.4833 \\
GeoSaddle fixed & 0.1203 & 0.0926 & 0.2617 \\
GeoSaddle final & 0.0981 & 0.0636 & 0.2177 \\
\midrule\multicolumn{4}{@{}l}{\textit{Network and paired outcomes}} \\
\multicolumn{2}{@{}l}{Formal A--B networks} & \multicolumn{2}{r@{}}{94/100} \\
\multicolumn{2}{@{}l}{Strict saddle-connected chains} & \multicolumn{2}{r@{}}{92/100} \\
\multicolumn{2}{@{}l}{Lower path RMS than BVSE or no detected difference} & \multicolumn{2}{r@{}}{81/92} \\
\multicolumn{2}{@{}l}{Closer fixed / final saddle} & \multicolumn{2}{r@{}}{64/92, 70/92} \\
\multicolumn{2}{@{}l}{DFT-path coverage at 0.20~\AA} & \multicolumn{2}{r@{}}{88.56\%} \\
\multicolumn{2}{@{}l}{One / two / three-saddle chains} & \multicolumn{2}{r@{}}{71 / 19 / 2} \\
\bottomrule
\end{tabularx}
\par\vspace{3pt}
\begin{minipage}{\columnwidth}
\footnotesize\raggedright
The whole-path RMS is computed from 21 arc-length samples after endpoint-compatible periodic symmetry alignment. For GeoSaddle relative to BVSE, the paired block-bootstrap interval for the mean squared distance difference lay below zero for 51 paths and above zero for 11. For 30 paths it included zero (Supplementary Sec.~\ref{si:evaluation}). GeoSaddle fixed and GeoSaddle final denote the selected saddles before and after local-host refinement. The final saddle set comprises 91 accepted local-host refinements and one retained fixed-host saddle. ``Closer saddle'' counts the cases in which the GeoSaddle saddle lies closer to the DFT path than the BVSE proxy. Over all 100 cases, the BVSE whole-path RMS has a mean, median and P90 of 0.1991, 0.1470 and 0.4358~\AA, and the BVSE saddle proxy has corresponding values of 0.2021, 0.1417 and 0.4886~\AA.
\end{minipage}
\end{table}

\begin{figure*}[!tbp]
\centering
\includegraphics[width=.96\textwidth]{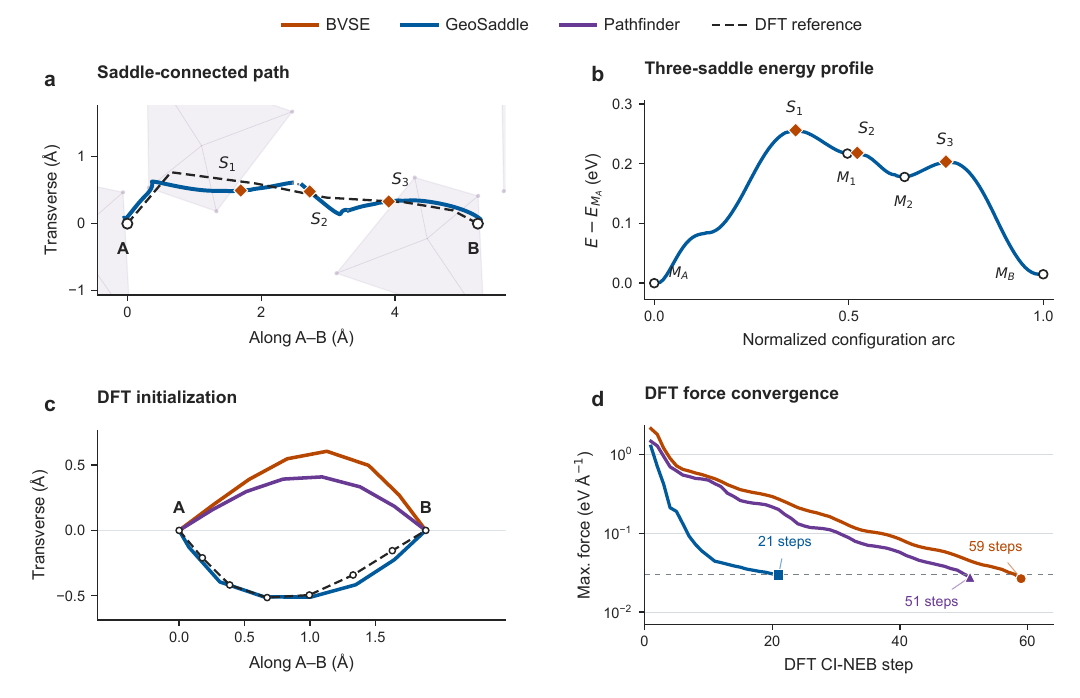}
\caption{\textbf{Saddle-connected pathway assembly and DFT NEB refinement.}
(a,b) Three-saddle GeoSaddle path in \ce{Li7Nb8P32O104} (\texttt{mp-684534}) and its fixed-host MatPES energy profile relative to $M_A$. Diamonds mark saddles and open circles mark minima. The dashed curve in (a) is the DFT reference. 
(c,d) Initial Li paths and DFT CI-NEB force convergence for \ce{Li5BO4} (\texttt{mp-768967}, 79 atoms). The dashed curve in (c) is the converged DFT path from the GeoSaddle guess. The dashed line in (d) marks the 0.03~eV\,\AA$^{-1}$ force threshold.}
\label{fig:pathway-neb}
\label{fig:li5bo4-dft-neb}
\end{figure*}

As two-sided relaxation ends in the actual energy basins adjacent to each saddle, the assembled graph resolves multi-step pathways without prior specification of intermediate sites. In 21 of the 92 recovered pathways, the $A$--$B$ chain passed through one or two intermediate minima (Table~\ref{tab:blind-main}). For example, in \ce{Li7Nb8P32O104} (Fig.~\ref{fig:pathway-neb}a,b), three verified saddles linked the endpoints through two intermediate basins in the sequence $M_A\to S_1\to M_1\to S_2\to M_2\to S_3\to M_B$.

To test whether GeoSaddle transfers beyond Li hosts, we applied the same workflow, with no parameters changed, to six Na hops in Fe, Mn and Ni olivine phosphates~\cite{petersen2026limitations}. These hops cover the dilute and single-vacancy compositions of 1 and 23 Na per 24 transition-metal ions, respectively. GeoSaddle recovered a strict single-saddle $A$--$B$ connection in all six cases, with mean and maximum saddle-to-path distances of 0.045 and 0.087~\AA\ (Supplementary Table~\ref{tab:na-main}). We also tested a Na1--Na2 hop in a single-vacancy \ce{Na4MnCr(PO4)3} NASICON cell using the same fixed-host workflow and archived DFT endpoints~\cite{behara2026nasicons}. GeoSaddle recovered a strict single-saddle connection with a whole-path RMS of 0.027~\AA\ and a saddle-to-path distance of 0.012~\AA\ relative to the DFT-NEB path with seven internal images.

\subsection{NEB refinement}
\label{sec:results-neb}

\begin{table}[!tbp]
\centering\small
\setlength{\tabcolsep}{4pt}\renewcommand{\arraystretch}{1.08}
\caption{\textbf{NEB comparison of three MEP guesses.} The comparison uses the same 15 LiTraj paths for all three initializations. Input RMS is measured in the shared LiTraj endpoint frame. Final Li-path RMS and NEB steps are from the MatPES-relaxed endpoint runs. All bands have 21 images.}
\label{tab:neb-main}
\begin{tabularx}{\columnwidth}{@{}Yrrr@{}}
\toprule
Quantity & BVSE & GeoSaddle & Pathfinder \\
\midrule
Mean initial Li-path RMS (\AA) & 0.252 & 0.089 & 0.222 \\
Mean final Li-path RMS (\AA) & 0.079 & 0.084 & 0.143 \\
Mean NEB steps & 98.9 & 70.9 & 101.9 \\
\bottomrule
\end{tabularx}
\end{table}

A continuous GeoSaddle chain joins the verified saddles and their descent paths into a complete MEP guess for climbing-image NEB (CI-NEB) using the \texttt{TensorNet-PES-MatPES-PBE-2025.2} FP. We compared the GeoSaddle guesses with BVSE and charge-density Pathfinder~\cite{rong2016pathfinder} guesses on 15 LiTraj paths for which the MatPES-relaxed endpoints passed the basin checks and all three initializations yielded a resolved climbing-image saddle (Table~\ref{tab:neb-main}). The GeoSaddle guesses started closest to the DFT reference, with a mean input RMS of 0.089~\AA\ compared with 0.252~\AA\ for BVSE and 0.222~\AA\ for Pathfinder. With the shared MatPES-relaxed endpoints, the GeoSaddle guess required a mean of 70.9 NEB steps, compared with 98.9 for BVSE and 101.9 for Pathfinder, reductions of 28\% and 30\%, respectively.

For the MatPES comparison, we relaxed all atoms in the LiTraj endpoint structures at fixed cell with a force target of 0.01~eV\,\AA$^{-1}$. We then reanchored all three initial paths to the same pair of relaxed endpoints and ran NEB. Across the 30 endpoints, the median Li displacement from the DFT position was 0.041~\AA, and the median framework RMS displacement was 0.038~\AA. The mean final Li-path RMS deviations from DFT were 0.084, 0.079 and 0.143~\AA\ for GeoSaddle, BVSE and Pathfinder, respectively. The forward-barrier mean absolute error relative to DFT was 0.066~eV for each initializer (Supplementary Table~\ref{tab:si-barriers}); the choice of initializer therefore changes the cost of NEB convergence but not the converged barrier.

We also performed DFT CI-NEB for \ce{Li5BO4}, one of these cases, in a single-vacancy cell with all 79 atoms relaxed (Fig.~\ref{fig:pathway-neb}c,d). The GeoSaddle guess started close to the converged DFT pathway, with a symmetric path RMS of 0.069~\AA. In contrast, the BVSE and Pathfinder guesses passed on the opposite side of the endpoint axis, with RMS values of 0.620 and 0.561~\AA, respectively. The GeoSaddle-initialized band reached the force criterion of 0.03~eV\,\AA$^{-1}$ in 21 ionic steps. This is less than half the number of steps required with the BVSE (59 steps) and Pathfinder (51 steps) guesses. All three bands converged to the same MEP, with barriers of 0.155--0.157~eV and final positions of the migrating Li that agree to within 0.050~\AA.

\subsection{Reconstruction of periodic defect mechanisms}
\label{sec:results-mechanisms}

\begin{figure*}[!tbp]
\centering
\includegraphics[width=.96\textwidth]{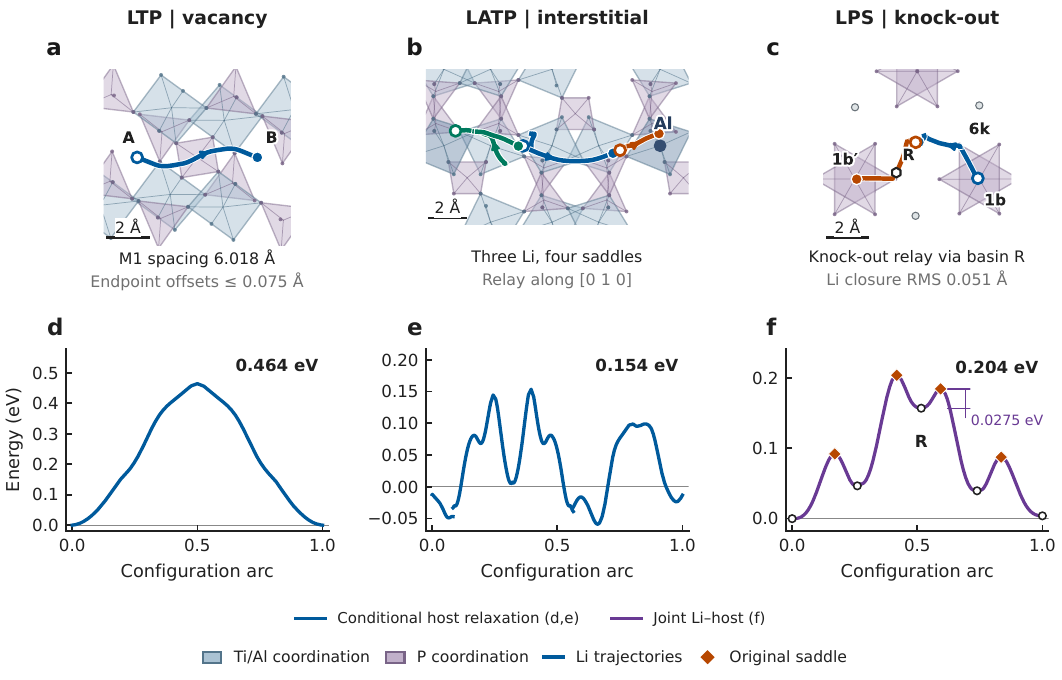}
\caption{\textbf{Defect relay mechanisms and framework accommodation in LTP, LATP and LPS.}
(a--c) Local coordination environments and Li trajectories for the M1--M2--M1 vacancy hop in LTP (a), the four-step three-Li split-interstitial relay near Al in LATP at $x=0.125$ (b) and the planar knock-out exchange through basin $R$ in LPS (c). The panels report the LTP endpoint offsets from the M1 sites, the LATP repeat direction in the reported cell setting and the LPS Li occupancy-closure RMS after one parent-cell $b$ translation. Colors distinguish Li identities, and open and filled circles mark initial and final positions (scale bars, 2~\AA).
(d,e) Conditional host-relaxed MatPES profiles along the normalized cumulative Li-configuration arc for LTP (d) and LATP (e). The host was relaxed at fixed Li coordinates, and energies are relative to the corresponding host-relaxed initial basin.
(f) LPS MatPES profile in the joint active space of six Li and 49 P/S atoms (purple), plotted against its normalized configuration arc and referenced to the initial minimum. Open circles mark minima, diamonds mark saddles and the bracket indicates the onward barrier from the shallow intermediate $R$.}
\label{fig:defects}
\end{figure*}

Macroscopic ionic transport requires percolating migration events. GeoSaddle follows such extended periodic mechanisms by starting new saddle searches from the relaxed product configurations of accepted events and matching atomic occupancies across periodic images (Sec.~\ref{sec:exploration}). We tested this capability using motif-guided continuation, in which a localized neighborhood of a vacancy or interstitial is activated and the search continues from each relaxed product until the mechanism closes periodically. The tests cover three model Li conductors with increasingly complex active spaces (Fig.~\ref{fig:defects}): single-ion vacancy hopping in NASICON-type LTP, a three-ion split-interstitial relay in Al-substituted LATP and a planar knock-out exchange coupled to polyhedral distortion in LPS. The reference studies used different exchange--correlation functionals, Hubbard-$U$ settings and defect-charge protocols, whereas the GeoSaddle searches were performed on MatPES in constrained active spaces. Our comparison therefore focuses on the reproduction of the mechanisms and the pathway geometry. The reference protocols are summarized in Supplementary Table~\ref{tab:literature-protocols}.

In NASICON-type LTP, each single-Li hop moved the M1 vacancy to the site vacated by the hopping Li in the relaxed product. Continuing the search from this product closed a two-event periodic relay along $[0,-1,0]$ (Fig.~\ref{fig:defects}a and Supplementary Table~\ref{tab:si-networks}). The reconstructed trajectory reproduced the M1--M2--M1 channel geometry of NASICON~\cite{arjmandi2019latp}, with relaxed endpoints within 0.075~\AA\ of the ideal M1 sites, which are 6.018~\AA\ apart. Conditional host relaxation, in which the non-Li framework is relaxed at fixed Li coordinates, gave a MatPES peak of 0.464~eV (Fig.~\ref{fig:defects}d). Published barriers for the corresponding M1--M2--M1 vacancy hop in LTP are 0.40~eV from PBE NEB~\cite{lu2017ltp} and 0.48~eV from PBEsol+$U$ CI-NEB~\cite{arjmandi2019latp}.

In LATP at $x=0.125$, which we modeled as \ce{Li9AlTi15P24O96}, Al substitution introduces excess Li, and the entry of an interstitial Li is coupled to the motion of neighboring Li ions. Three-Li saddle searches with product continuation mapped a local transition network of 45 states and reconstructed a four-saddle relay along $[0,1,0]$ that follows the reported split-interstitial mechanism (Fig.~\ref{fig:defects}b)~\cite{arjmandi2019latp,pfalzgraf2021latp}. In this relay, one ion of the split pair near Al crosses M2 to form a new pair at the next M1 site, while its original partner returns toward M1. The conditional host-relaxed MatPES profile peaks at 0.154~eV relative to the host-relaxed initial basin (Fig.~\ref{fig:defects}e).

For the planar knock-out exchange in LPS~\cite{stamminger2020lps}, we verified the mechanism in a coupled space of 165 coordinates comprising six Li and 49 P/S atoms. This verification resolved five minima and four saddles within an energy span of 0.204~eV, including a shallow relay intermediate $R$ with an onward barrier of 0.0275~eV (Fig.~\ref{fig:defects}c,f). The incoming Li from a $1b$ site replaced the lattice Li at $2d$ and settled within 0.162~\AA\ of the lattice site, and the displaced Li reached the next $1b'$ site within 0.085~\AA. Over the full cell, all 33 Li positions closed with an RMS residual of 0.051~\AA\ after one translation along the parent-cell $b$ vector.

\begin{figure*}[!tbp]
\centering
\includegraphics[width=.94\textwidth]{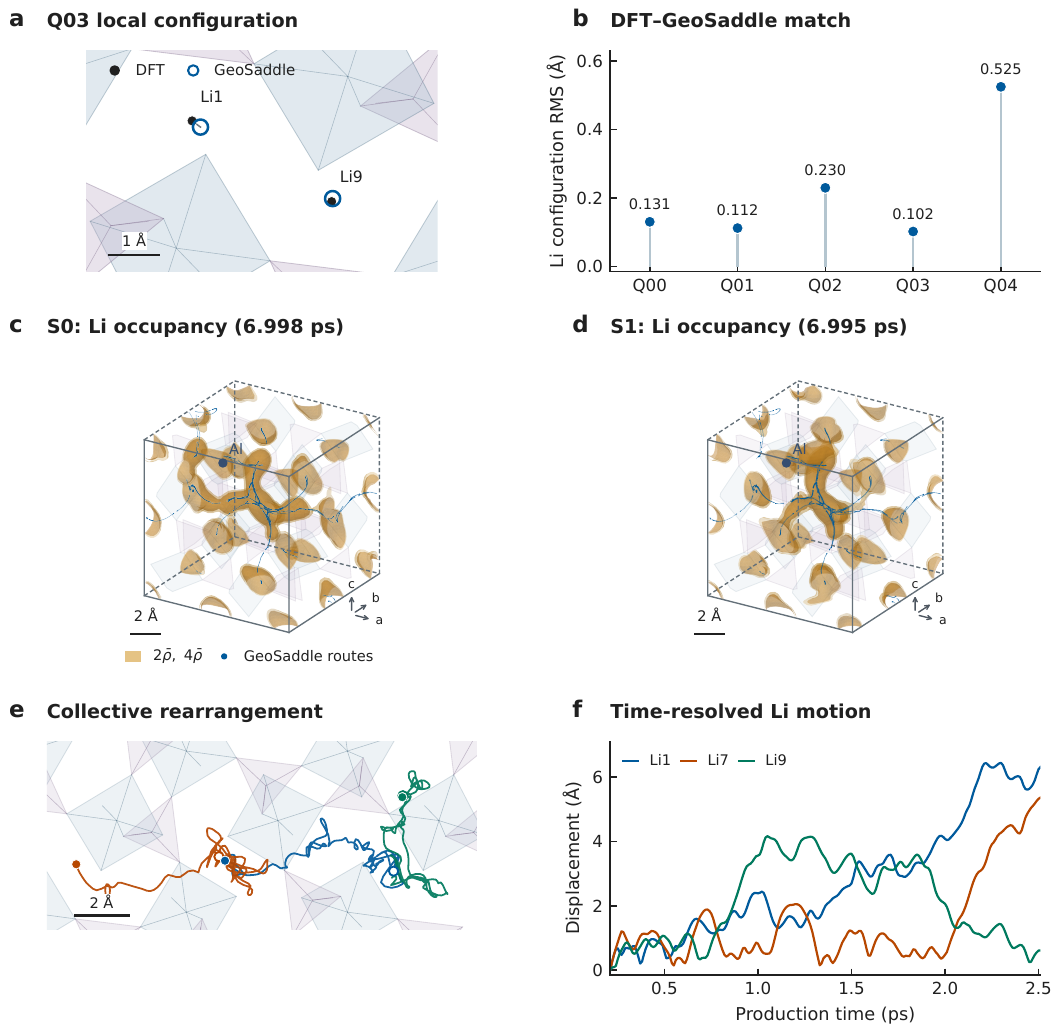}
\caption{\textbf{First-principles structural and dynamical validation in LATP.}
(a) Local Li positions in full-ion DFT-quenched \textit{ab initio} molecular dynamics (AIMD) snapshot Q03 (filled black) and its closest GeoSaddle network state (open blue), with Ti/Al--O (blue-gray) and P--O (violet-gray) polyhedra.
(b) Minimum nine-Li RMS correspondence between five DFT-quenched AIMD snapshots and the 45-state GeoSaddle network.
(c,d) Three-dimensional Li probability-density isosurfaces at $2\bar\rho$ and $4\bar\rho$ from two independent 900~K AIMD trajectories (S0, 6.998~ps and S1, 6.995~ps) overlaid with GeoSaddle route samples (blue points).
(e,f) Cooperative Li1, Li7 and Li9 trajectories over 0.210--2.510~ps in S1, with open and filled markers denoting start and end positions (e), and corresponding displacements from the event start (f).}
\label{fig:latp-aimd}
\end{figure*}

The LATP reconstruction kept the host framework fixed during the saddle searches. We therefore used \textit{ab initio} molecular dynamics (AIMD) with PBEsol+$U$ at 900~K to test whether the GeoSaddle configurations and channels remain accessible when all 145 atoms move thermally (Sec.~\ref{sec:evaluation}). We ran two trajectories of about 7~ps each. The first started from a DFT-relaxed GeoSaddle state, S0, and the second from a state S1 that was generated by applying one three-Li GeoSaddle displacement to S0. We then quenched five snapshots to full-ion DFT minima. Snapshot Q00 returned to the S0 starting state, with a nine-Li RMS deviation of 0.021~\AA, whereas Q01--Q04 relaxed to configurations 0.57--1.19~\AA\ from the starting states of their trajectories (Supplementary Table~\ref{tab:latp-quenches}). Three of these four displaced states, Q01--Q03, matched off-chain branches of the 45-state GeoSaddle network, within a nine-Li RMS of 0.102--0.230~\AA\ (Fig.~\ref{fig:latp-aimd}a,b). The fourth state, Q04, lay 0.525~\AA\ from its closest network state and remained unmatched.

In both 900~K runs, the Li probability-density isosurfaces followed the fixed-host GeoSaddle channel network (Fig.~\ref{fig:latp-aimd}c,d). During the 0.210--2.510~ps collective rearrangement in S1, Li1 and Li7 moved within a 0.276~\AA\ pooled median distance of the GeoSaddle route samples while Li9 participated transiently (Fig.~\ref{fig:latp-aimd}e,f).

\subsection{Collective pathway exploration}
\label{sec:results-lgps}

\begin{figure*}[!tbp]
\centering
\includegraphics[width=.96\textwidth]{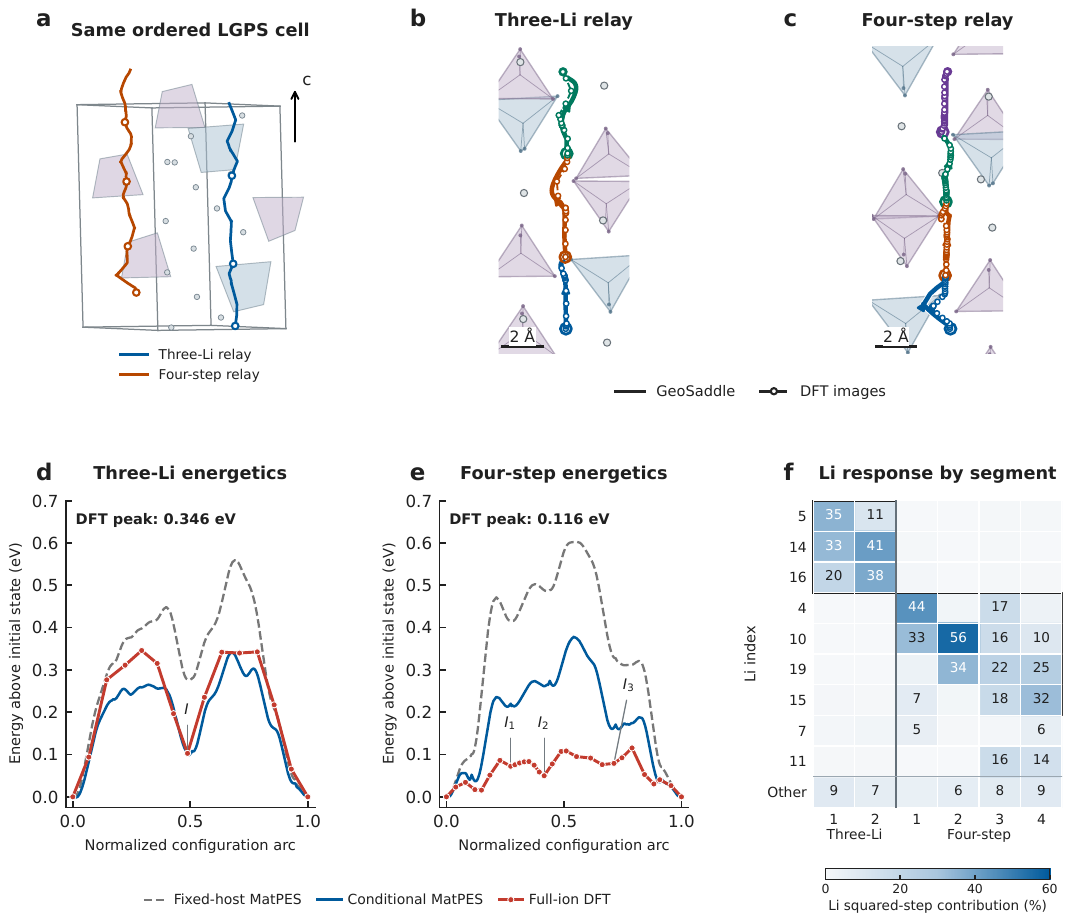}
\caption{\textbf{Two LGPS relay cycles under MatPES screening and full-ion DFT refinement.}
(a) Both DFT-refined cycles within the same ordered 50-atom LGPS cell. Blue and orange show the distinct $c$-axis channels of the three-Li and four-step relays, each transporting one net Li by $+c$ from the shared initial state $A$.
(b,c) GeoSaddle polylines (solid) and full-ion DFT CI-NEB images (open circles, dashed joins) colored by persistent Li identity, shown with \ce{GeS4}/\ce{PS4} polyhedra (scale bars, 2~\AA).
(d,e) Energy profiles referenced to each method's initial state $A$ along the normalized relay-Li configuration arc for the three-Li (d) and four-step (e) cycles. Curves show fixed-host MatPES (gray dashed), conditional host-relaxed MatPES (blue solid) and full-ion DFT CI-NEB (red circles).
(f) Per-Li contributions (\%) to summed squared image-to-image displacements across the two three-Li and four four-step DFT bands (zero-based Li indices). Outlined boxes mark designated relay ions, and values $\ge 5\%$ are labeled.}
\label{fig:lgps-dft-validation}
\end{figure*}

Unlike isolated point-defect hops, collective transport in a densely occupied superionic sublattice has no predefined active defect or endpoint pair. For LGPS, we therefore used sublattice exploration (Sec.~\ref{sec:exploration}), in which all mobile ions are active and the network is expanded from a single starting state. In the ordered 50-atom \ce{Li20Ge2P4S24} cell, we started from one relaxed structure, $A$, with the 30-atom Ge/P/S framework fixed and all 20 Li ions active, which gives 60 active coordinates. We supplied neither endpoints nor labels for the mobile ions. Graph expansion from state $A$ mapped a local transition network of 132 states, 230 verified transitions and 12 cycles with nonzero winding, which carry net Li transport across two independent crystal directions and give a transport rank of $r=2$ (Supplementary Table~\ref{tab:si-networks}). From this network, we examined two multi-step relays in distinct $c$-axis channels. Each relay closes on a Li permutation of $A$ with a net transport of one Li along $+c$. The unbiased queue expansion first closed a two-saddle, three-Li relay cycle. Targeted four-Li proposals and continuation within the same graph then recovered a four-step, four-Li relay, which belongs to the family of concerted channel mechanisms reported previously~\cite{mo2012lgps,he2017concerted}. We refer to these as the three-Li relay and the four-step relay, respectively. Their fixed-host and conditional host-relaxed MatPES profiles, together with those of a single-saddle four-Li event, are given in Supplementary Fig.~\ref{fig:si-lgps-matpes} and Supplementary Table~\ref{tab:lgps-main}. We then refined both relays by full-ion PBE DFT CI-NEB, with all 50 atoms relaxed, between the DFT-relaxed state $A$ and the full-ion-relaxed intermediate junctions (Fig.~\ref{fig:lgps-dft-validation}).

Both cycles remained connected under full-ion DFT refinement. They preserved the GeoSaddle transition sequence, the monotonic progression of the images and the \ce{GeS4}/\ce{PS4} coordination (Fig.~\ref{fig:lgps-dft-validation}a--c). Over the six bands, the RMS deviations of the relay Li ions in the internal images from the GeoSaddle polylines ranged from 0.14 to 0.34~\AA\ (Supplementary Table~\ref{tab:si-lgps-six}). The designated relay ions accounted for 70--93\% of the squared Li displacements, and the transient motion of two neighboring Li ions accounted for most of the remainder (Fig.~\ref{fig:lgps-dft-validation}f).

On the full-ion DFT surface, the three-Li relay configurations remained near crystallographic sites and shifted by only 0.08--0.13~\AA\ all-Li RMS upon relaxation, which left the DFT peak at 0.346~eV above $A$, close to its conditional host-relaxed MatPES peak of 0.342~eV (Fig.~\ref{fig:lgps-dft-validation}d). Along the four-step channel, where several relay ions lie 0.80--1.18~\AA\ from the nearest experimental Li sites (Supplementary Fig.~\ref{fig:si-lgps-registry}), coupled Li--framework relaxation shifted the three shallow intermediates by 0.27--0.35~\AA\ all-Li RMS to 0.050--0.079~eV above $A$. Full-ion DFT refinement yielded an overall peak of 0.116~eV, compared with 0.378~eV on conditional host-relaxed MatPES, and preserved all three intermediate basins and their cyclic Li-exchange sequence (Fig.~\ref{fig:lgps-dft-validation}e). 

\begin{figure*}[!tbp]
\centering
\includegraphics[width=.96\textwidth]{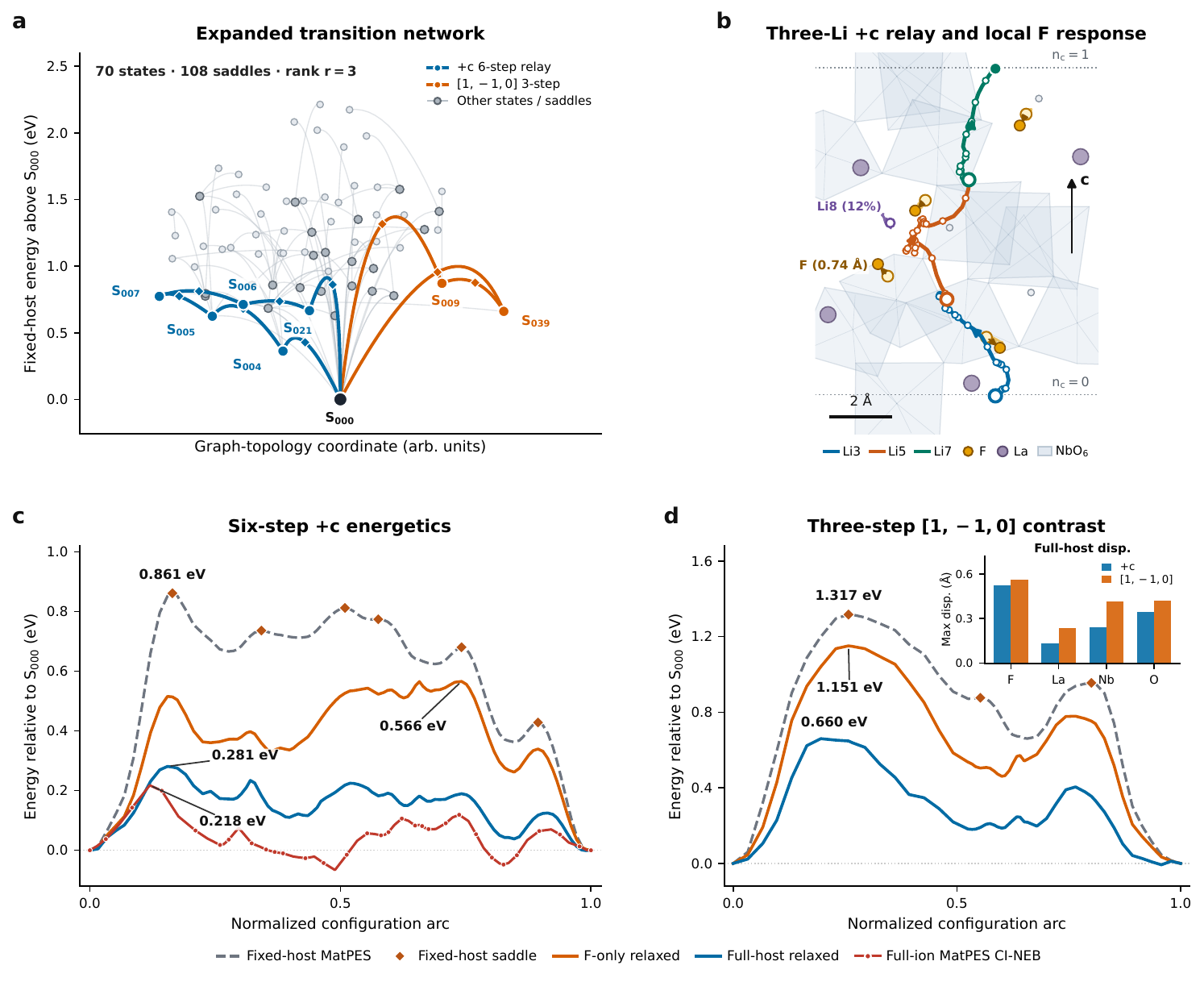}
\caption{\textbf{Expanded transition network, three-Li relay and host-relaxation energetics in LLNOF (\ce{Li9La5Nb16O48F8}).}
(a) Fixed-host transition network of 70 states and 108 verified saddles ($r=3$) plotted by state energy relative to $S_{000}$ against a graph-topology coordinate. Highlighted curves and diamonds trace the six-step $[0,0,1]$ relay (blue) and the three-step $[1,-1,0]$ relay (orange).
(b) Full-ion MatPES CI-NEB trajectory along $+c$, with La ions shown at their initial positions. Colors track Li3, Li5, and Li7, with Li8 shown in dashed purple. Open and filled circles mark initial and final Li positions. Amber arrows show maximum F displacements (up to 0.74~\AA). Scale bar, 2~\AA.
(c) Fixed-host, F-only relaxed, full-host relaxed and full-ion MatPES CI-NEB energy profiles along the normalized Li-configuration arc of the six-step $[0,0,1]$ relay, referenced to each protocol's initial state $S_{000}$.
(d) Corresponding conditional profiles along the three-step $[1,-1,0]$ relay, and the inset compares maximum host atomic displacements under conditional full-host relaxation (fixed Li) between the two channels.}
\label{fig:llnof-mechanisms}
\end{figure*}

We extended the same sublattice exploration to the pyrochlore oxyfluoride \ce{Li_{2-x}La_{(1+x)/3}Nb2O6F} (LLNOF)~\cite{aimi2024llnof,jalem2024llnof}. In an ordered 86-atom \ce{Li9La5Nb16O48F8} cell at $x=0.875$, we started from one relaxed structure, $S_{000}$, with the 77-atom La/Nb/O/F host fixed and all nine Li ions active, which gives 27 active coordinates. GeoSaddle mapped a local transition network of 70 states and 108 verified transitions spanning all three independent crystal directions, which gives a transport rank of $r=3$ (Fig.~\ref{fig:llnof-mechanisms}a). Within this network, the lowest-barrier cycle is a six-step, three-Li relay with a net transport of one Li along $+c$ (Fig.~\ref{fig:llnof-mechanisms}b). Along the same fixed Li trajectory, relaxation of the F ions alone lowered the MatPES profile peak from 0.861 to 0.566~eV, and full-host relaxation reduced it further to 0.281~eV (Fig.~\ref{fig:llnof-mechanisms}c). Because the ordered La arrangement breaks the cubic symmetry of $Fd\bar{3}m$ LLNOF, channel barriers are anisotropic in this cell: a three-step relay along $[1,-1,0]$ started at a fixed-host peak of 1.317~eV, dropped to 1.151~eV under F-only relaxation and remained at 0.660~eV after full-host relaxation as nearby La and Nb atoms displaced more strongly (Fig.~\ref{fig:llnof-mechanisms}d). Under full-ion MatPES CI-NEB refinement of the $+c$ relay, all six segments converged and preserved the periodic winding, with a peak of 0.218~eV above $S_{000}$, an overall energy span of 0.284~eV and local F displacements of up to 0.740~\AA. For context at nearby solid-solution compositions, bulk impedance measurements at $x=0.75$ report activation energies of 0.22~eV below 200~K and 0.14~eV above 200~K~\cite[Fig.~1b]{kuwata2025llnof}, and PBE AIMD simulations at $x=0.6875$ yield 0.19--0.21~eV~\cite{jalem2024llnof}. These profiles separate the energetic contribution of local fluoride relaxation from that of the remaining framework in this Li/La/vacancy configuration.


\FloatBarrier
\section{Discussion and conclusion}
\label{sec:discussion-conclusion}

In summary, we have developed GeoSaddle, which identifies ion migration paths together with their transition states and assembles them into periodic transport networks on a PES.

GeoSaddle constructs each migration path by locating transition states and relaxing downhill from them to the adjacent minima, and the resulting transition states and pathways lie close to the DFT references across Li and Na host structures. A geometric analysis of the host crystal identifies candidate positions and migration directions within accessible channels, without requiring precomputed PES grids. This geometric initialization concentrates model evaluations on local saddle searches within the channels. For ten long, curved events, it required 12.7\% of the model evaluations used by an ellipsoidal PES-grid control. On LiTraj-100, whose reference DFT NEB paths were originally initialized from BVSE guesses, GeoSaddle recovered verified pathways without reference interior images. These pathways matched the converged DFT paths more closely, on average, than the BVSE guesses themselves.

The same construction supports both the validation of candidate mechanisms and the discovery of pathways beyond single-step hops. Because two-sided relaxation terminates in the actual energy basins adjacent to each saddle, graph assembly resolves the intermediate minima of multi-step pathways between fixed endpoints and retains metastable states that lie off the selected pathway. Across unit cells, state-to-state continuation from the actual relaxed products and periodic occupancy matching extended local transitions into propagating vacancy, split-interstitial and knock-out relays in LTP, LATP and LPS, respectively. In LATP, the resulting network also captured thermally accessible off-chain states, which matched configurations quenched from 900~K AIMD. In LGPS, the same procedure identified distinct three-Li and four-Li relay cycles from a shared initial state, each with net Li transport along $+c$. In LLNOF, it resolved a six-step, three-Li relay and isolated the energetic contribution of local fluoride relaxation. The sensitivity of the calculated relay to fluoride relaxation provides a pathway-level complement to the directional F off-centering observed in LLNOF single crystals~\cite{yajima2026llnof}.

The saddle, minimum and pathway structures obtained on an FP transfer directly to first-principles refinement. Small deviations between the FP and DFT, together with full-lattice relaxation, have little effect when migrating ions remain near crystallographic sites, as in the three-Li relay of LGPS, but can shift or flatten shallow off-site basins and adjust barrier heights when ion motion couples strongly to framework relaxation, as in the four-step relay of LGPS. Even so, the channel connectivity and transition sequence identified by GeoSaddle remain intact: all six full-ion DFT CI-NEB bands in LGPS converged smoothly along the GeoSaddle sequence without disconnection. With shared MatPES-relaxed endpoints, GeoSaddle reduced the mean number of NEB steps by 28--30\% relative to BVSE and Pathfinder, and in \ce{Li5BO4}, where heuristic guesses deviated from the transition valley, reached the DFT CI-NEB force threshold in 21 ionic steps, compared with 59 for BVSE and 51 for Pathfinder, without altering the converged MEP or barrier.

Together with FPs, GeoSaddle bridges static band interpolation and dynamical trajectory sampling for crystal ion migration from the host structure alone. Unlike two-endpoint NEB, which optimizes a band along a predetermined path topology between fixed initial and final states, GeoSaddle propagates state by state from actual relaxed products to explore branching intermediates and concerted multi-ion cycles. At the same time, whereas high-temperature AIMD samples thermally accessible channels stochastically over picosecond windows without isolating zero-temperature transition states, GeoSaddle directly locates and verifies the underlying index-one saddles and minimum-to-minimum connections at a fraction of the computational cost. The verified saddle and intermediate crystal structures generated by GeoSaddle provide connected starting configurations for targeted DFT optimization across solid electrolytes and electrodes, as well as transition-region structures for first-principles evaluation and interatomic potential development. We expect the workflow to extend to other mobile species and to disordered host frameworks, and the open-source GeoSaddle code enables its use with any PES that supplies energies and forces.

\section{Methods}
\label{sec:methods}

GeoSaddle first analyzes the channel geometry of the host crystal without any PES evaluations, and then uses local searches on a PES to locate and verify stationary points (Fig.~\ref{fig:gpa-architecture}). Columns of the lattice matrix $\mathbf L$ store the crystal lattice vectors, and lifted jump endpoints represent the specified periodic images of an ion hop. Energy and force evaluations act on a chosen active coordinate vector $\mathbf q$ with the remaining coordinates $\mathbf R_{\mathrm{fix}}$ and cell $\mathbf L$ held fixed, which defines the constrained surface
$U_{\mathcal A}(\mathbf q)=U(\mathbf q,\mathbf R_{\mathrm{fix}};\mathbf L)$
for movable-atom set $\mathcal A$. Different choices of $\mathcal A$ and proposal rules adapt the same workflow across single-ion endpoint pathways and vacancy, interstitial or collective multi-ion exploration.

\subsection{Geometric initialization of saddle searches}
\label{sec:geometry}

To quantify local free space at position $\mathbf x$, we remove the mobile ions of interest from the host obstacle set $\mathcal O$ and compute the periodic clearance to host spheres at $\mathbf R_j$ across integer lattice shifts $\mathbf n\in\mathbb Z^3$:
\begin{equation}
c(\mathbf x)=
\min_{j\in\mathcal O,\,\mathbf n\in\mathbb Z^3}
\bigl[\|\mathbf x-\mathbf R_j-\mathbf L\mathbf n\|
-\rho_j^{\mathrm{eff}}\bigr],
\label{eq:clearance}
\end{equation}
where $\rho_j^{\mathrm{eff}}$ combines tabulated ionic radii with a bond-valence-dependent radial allowance (Supplementary Sec.~\ref{si:geometry}).

Pointwise clearance alone cannot distinguish a through-going transport channel from an isolated pocket or dead-end cavity. We therefore discretize positive-clearance space onto a lifted periodic graph with conservative segment clearance capacities $w_e$ on edges $e$ and evaluate bottleneck connectivity to both lifted endpoints $A$ and $B$ via the widest-path capacity $W_A(v)$ and endpoint-conditioned aperture $\alpha_{AB}(v)$:
\begin{equation}
\begin{aligned}
W_A(v)&=\max_{\pi:A\leadsto v}\min_{e\in\pi}w_e,\\
\alpha_{AB}(v)&=\min\{W_A(v),W_B(v)\}.
\end{aligned}
\label{eq:aperture}
\end{equation}
Here $\pi:A\leadsto v$ denotes a path from $A$ to vertex $v$, including continuous endpoint connectors. The set $\mathcal T_a=\{v:\alpha_{AB}(v)\ge a\}$ contains vertices reachable from both endpoints through bottlenecks of width at least $a$. At each threshold $a$, we find the shortest admissible path $\gamma_a$ from $A$ to $B$ with edge lengths $\ell_e$:
\begin{equation}
\gamma_a\in\argmin_{\pi:A\leadsto B}\sum_{e\in\pi}\ell_e
\quad\text{s.t.}\quad w_e\ge a\;\;\forall e\in\pi.
\label{eq:route}
\end{equation}
We block previously sampled bottlenecks to find alternative routes, remove duplicates and retain centerlines $\{\gamma_k\}$ that meet the minimum probe clearance.
We then sample each piecewise-linear centerline $\gamma_k(s)$ at interior chord-length fractions $s_j\in(0,1)$, together with the direct $A$--$B$ midpoint, to obtain initial positions and unit tangent modes without PES evaluations (Supplementary Algorithm~\ref{alg:geometry}):
\begin{equation}
\mathbf q^{(0)}_{kj}=\gamma_k(s_j),\qquad
\mathbf v^{(0)}_{kj}=\frac{\dot\gamma_k(s_j)}{\|\dot\gamma_k(s_j)\|}.
\label{eq:seed}
\end{equation}

\subsection{Construction of saddle-connected pathways}
\label{sec:path-construction}

Each directional seed $(\mathbf q^{(0)},\mathbf v^{(0)})$ initializes a minimum-mode dimer search~\cite{henkelman1999dimer} on $U_{\mathcal A}$ that rotates toward the lowest-curvature mode and translates uphill toward an index-one saddle. Stationary candidates are evaluated from the maximum active-atom force $F_{\max}$ and the negative-eigenvalue count $I_{\varepsilon_\lambda}(H)$ of the symmetrized active-space finite-difference Hessian $H$:
\begin{equation}
F_{\max}=\max_{i\in\mathcal A}\|\nabla_{\mathbf r_i}U\|,\qquad
I_{\varepsilon_\lambda}(H)=\#\{\lambda_i(H)<-\varepsilon_\lambda\},
\label{eq:classification}
\end{equation}
where $\varepsilon_\lambda>0$ is the curvature tolerance. An index-one candidate must meet the force convergence criterion and have $I_{\varepsilon_\lambda}(H)=1$ in the active space. Strict single-ion endpoint tests also require the energy to decrease on both sides of the saddle when displaced by $\pm 0.10$~\AA\ along the unstable eigenvector $\mathbf v_-$. For collective crystal searches, we verify connectivity through two-sided descent to adjacent minima so that curved valleys are accommodated (Supplementary Sec.~\ref{si:verification}).

To identify the basins connected by an eligible saddle $\mathbf q_S$, we displace $\mathbf q_S$ by $\pm\varepsilon_d\mathbf v_-$ with $\varepsilon_d>0$ and relax both configurations via the fast inertial relaxation engine (FIRE)~\cite{bitzek2006fire}, denoted $\mathcal R_{\mathcal A}$, within the same active space:
\begin{equation}
H_S\mathbf v_-=\lambda_-\mathbf v_-\;\;(\lambda_-<0),\qquad
\mathcal R_{\mathcal A}(\mathbf q_S\pm\varepsilon_d\mathbf v_-)\longrightarrow M_\pm.
\label{eq:descent}
\end{equation}
A verified elementary event comprises the index-one saddle $S$, its two terminal minima $M_\pm$ and both descent trajectories~\cite{vineyard1957,wales2002dps}. We merge periodically equivalent minima and saddles to form a finite sample of the local Morse--Smale complex on $U_{\mathcal A}$ (Supplementary Fig.~\ref{fig:morse-smale}). If $A$ and $B$ remain disconnected after primary seeding, adaptive recovery launches targeted radical-Voronoi, bridge and minimum-to-minimum string searches to discover intermediate basins and update the graph, and index-two string peaks are split into transverse index-one child searches (Supplementary Sec.~\ref{si:network}). On the strict endpoint graph, the representative pathway $P^\star$ along $M_A\to S_1\to M_1\to\cdots\to S_m\to M_B$ minimizes the highest saddle energy across connected chains $\mathcal P_{AB}$, with ties broken by edge count:
\begin{equation}
P^\star\in\argmin_{P\in\mathcal P_{AB}}\max_{S\in P}E(S),\qquad
\Delta E_{i\to j}=E(S_{ij})-E(M_i),
\label{eq:chain}
\end{equation}
with the highest-energy saddle along $P^\star$ retained as the representative transition state.

\subsection{Periodic network assembly and mechanism exploration}
\label{sec:exploration}

To follow mechanisms beyond a single prescribed jump, accepted product configurations supply updated source coordinates $\mathbf q_{\mathrm{src}}$ and site occupancies for subsequent proposals:
\begin{equation}
\mathbf q^{(0)}=\mathbf q_{\mathrm{src}}+\Delta\mathbf q,\qquad
\mathbf v^{(0)}=\mathbf m/\|\mathbf m\|,
\label{eq:collective-seed}
\end{equation}
where trial displacements $\Delta\mathbf q$ and initial modes $\mathbf m$ direct active ions along accessible GeoSaddle channels together with induced neighbor displacements.

Local transitions are organized into periodic networks by tracking atomic identities and lattice-image shifts relative to a persistent host-framework gauge. For each verified event $(M_-,S,M_+)$, the reversed $M_-$ descent trajectory, saddle $\mathbf q_S$ and $M_+$ descent trajectory are concatenated, unwrapped across periodic boundaries and shifted to remove uniform host drift to form a continuous local path $X_-\to S\to X_+$. Each path endpoint $X\in\{X_-,X_+\}$ is then matched against canonical registered states $R$ by a periodic linear assignment over species-preserving permutations $\Pi_{\mathrm{sp}}$:
\begin{equation}
(\pi,\{\mathbf m_i\})\in\argmin_{\pi\in\Pi_{\mathrm{sp}},\,\mathbf m_i\in\mathbb Z^3}
\sum_i \bigl\|\mathbf r_{\pi(i)}^X-\mathbf r_i^R-\mathbf L\mathbf m_i\bigr\|^p,
\label{eq:registration}
\end{equation}
with $p=1$ for LATP and $p=2$ for LPS, LGPS and LLNOF (Supplementary Sec.~\ref{si:network}). Endpoints within the species-wise coordinate and energy tolerances merge into existing graph nodes, whereas unmatched minima define new canonical states. Composition of the endpoint matches $(\pi_-,\mathbf m_-)$ at $R_-$ and $(\pi_+,\mathbf m_+)$ at $R_+$ assigns each oriented transition edge $e:R_-\to R_+$ a species-preserving permutation $\sigma_e=\pi_+^{-1}\circ\pi_-$ and integer lattice shifts $\mathbf n_{e,i}=\mathbf m_{+,\sigma_e(i)}-\mathbf m_{-,i}\in\mathbb Z^3$.

Along any multi-step walk $C=(e_1,\dots,e_m)$ on the state graph, sequential composition of $(\sigma_{e_k},\mathbf n_{e_k})$ relabels and translates the local segments $X_-^{(k)}\to S^{(k)}\to X_+^{(k)}$ into a continuous multi-event trajectory across unit cells. When $C$ closes on its initial canonical state $R_m=R_0$, the cycle permutation $\sigma_C=\sigma_{e_m}\circ\cdots\circ\sigma_{e_1}$ identifies the cyclic ion-exchange relay and separates relay ions that advance onto periodic copies of neighboring sites from ions that return to their initial sites, while the accumulated integer shifts $\mathbf n_i(C)$ over mobile ions $i\in\mathcal M$ define the net ionic winding $\mathbf w(C)$ and transport rank $r$:
\begin{equation}
\begin{gathered}
\mathbf w(C)=\sum_{i\in\mathcal M}\mathbf n_i(C),\\
r=\operatorname{rank}_{\mathbb R}
\{\mathbf w(C):C\text{ is a verified cycle}\}.
\end{gathered}
\label{eq:periodicity}
\end{equation}
Root-reachable cycles are audited on a spanning tree of the transition graph: a nonzero $\mathbf w(C)$ certifies net periodic ion transport, including single-edge transitions that map a state onto a permuted periodic image of itself, and $r$ counts the independent crystallographic transport directions spanned by verified cycles.

We operate this framework in two modes. First, \emph{motif-guided continuation} in LTP, LATP and LPS activates localized defect neighborhoods around a vacancy or interstitial to verify elementary rearrangements and continue from the relaxed product state until periodic closure is reached. Second, \emph{sublattice exploration} in LGPS and LLNOF (Supplementary Algorithm~\ref{alg:exploration}) activates all mobile Li ions (20 in LGPS and nine in LLNOF) in a state queue from a single crystal structure. It retains off-state saddles, launches ridge proposals toward disconnected basins and rebuilds local channel skeletons upon stagnation until verified cycles reach nonzero winding, supplemented by targeted multi-ion proposals and state continuations.

\subsection{Computational and evaluation procedures}
\label{sec:evaluation}

\paragraph{Model and active-space protocols.}
All FP calculations fixed the lattice vectors and used the pretrained \texttt{TensorNet-\allowbreak PES-\allowbreak MatPES-\allowbreak PBE-\allowbreak 2025.2} FP~\cite{kaplan2025matpes,simeon2023tensornet,ko2025matgl} with pymatgen~\cite{ong2013pymatgen} and the Atomic Simulation Environment (ASE)~\cite{larsen2017ase}. Single-ion endpoint searches moved only the three Cartesian coordinates of the migrating ion. In LiTraj-100, selected fixed-host saddles were further refined by local-host dimer searches that included non-Li framework atoms within 4.0~\AA, whereas the ten-case seed test and the Na hops were evaluated in the fixed-host space. For LTP, LATP, LGPS and LLNOF, conditional host relaxation optimized non-Li framework atoms at fixed Li coordinates relative to the identically constrained initial basin to quantify framework accommodation, with an additional F-only relaxation mask and full-ion MatPES CI-NEB refinement applied to LLNOF, whereas LPS was verified in a joint 165-coordinate space of six Li and 49 P/S atoms (Supplementary Table~\ref{tab:si-active-spaces} and Supplementary Sec.~\ref{si:refinement}).

\paragraph{Seed comparison.}
Across ten long, curved LiTraj-100 paths, we compared GeoSaddle channel seeds with local random displacements around relaxed endpoints and the ellipsoidal PES-grid control under identical single-ion dimer settings, strict saddle criteria and matched per-case budgets of 97 attempts in total. Cost counts all FP force calls across seed preparation, dimer translation/rotation and Hessian/probe verification. A near-path hit is a strict saddle within 0.20~\AA\ of the continuous DFT Li polyline evaluated post hoc (Supplementary Sec.~\ref{si:evaluation}).

\paragraph{Pathway recovery.}
Blind recovery was tested on LiTraj-100~\cite{dembitskiy2025litraj} with fixed hyperparameters and no access to reference interior images, together with six Na olivine-phosphate hops~\cite{petersen2025sodium,petersen2026limitations} and one Na hop in a NASICON~\cite{behara2026nasicons}. Network connectivity is reported over all cases, while paired geometric metrics evaluate the 92 strict LiTraj-100 chains relative to BVSE and the Na chains via saddle-to-polyline distance, 0.20~\AA\ DFT-path coverage and 21-point symmetry-aligned periodic nearest-polyline RMS deviation:
\begin{equation}
\mathrm{RMS}_{\mathrm{path}}=
\left[\frac1K\sum_{k=1}^{K}
d_{\mathrm{PBC}}(\mathbf x_k,\mathcal P_{\mathrm{DFT}})^2\right]^{1/2},
\label{eq:metric}
\end{equation}
where $d_{\mathrm{PBC}}(\mathbf x_k,\mathcal P_{\mathrm{DFT}})$ is the minimum periodic distance from each of the $K=21$ arc-length samples $\mathbf x_k$ to the continuous DFT polyline $\mathcal P_{\mathrm{DFT}}$. Auxiliary path-initialization and candidate pre-screening controls on an independent 98-event LiTraj cohort are provided in Supplementary Sec.~\ref{app:initialization}.

\paragraph{Matched NEB refinement.}
For the matched NEB comparison with verified double-vacancy DFT charge-density inputs, BVSE, GeoSaddle and charge-density Pathfinder~\cite{rong2016pathfinder} paths were resampled to 21 images with identical endpoints, linearly interpolated internal hosts and full-ion MatPES FIRE/CI-NEB settings with spring constant $k=0.1$~eV\,\AA$^{-2}$ and a 0.02~eV\,\AA$^{-1}$ climbing-image force criterion. Input RMS uses the original shared LiTraj endpoint frame. We relaxed all endpoint atoms on MatPES at fixed cell to a force target of 0.01~eV\,\AA$^{-1}$ and reanchored the initial bands before NEB. The main-text comparison uses the 15 cases with preserved endpoint basins and a resolved climbing-image saddle for all three initializations.

\paragraph{First-principles DFT CI-NEB and AIMD.}
First-principles DFT calculations were performed with the Vienna Ab initio Simulation Package (VASP)~\cite{kresse1996vasp} within the projector augmented-wave (PAW) formalism~\cite{blochl1994paw,kresse1999paw} at a 520~eV plane-wave cutoff and fixed cell (Supplementary Secs.~\ref{si:refinement} and~\ref{si:latp-aimd}). For the 79-atom \ce{Li5BO4} cell \texttt{mp-768967}, all three initializers were refined by full-ion Perdew--Burke--Ernzerhof (PBE)~\cite{perdew1996pbe} CI-NEB from shared DFT-relaxed endpoints with six internal images, a $\Gamma$-centered $3\times 3\times 3$ $k$-point mesh and a 0.03~eV\,\AA$^{-1}$ force threshold. For the 50-atom \ce{Li20Ge2P4S24} cell of LGPS, full-ion PBE relaxations on a $\Gamma$-centered $2\times 2\times 2$ $k$-point mesh to $f_{\max}<0.01$~eV\,\AA$^{-1}$ defined the common initial state and intermediate junctions for two six-image CI-NEB bands of the three-Li relay and four six-image bands of the four-step relay, each converged to $f_{\max}<0.05$~eV\,\AA$^{-1}$. For the 145-atom \ce{Li9AlTi15P24O96} cell of LATP, spin-polarized $\Gamma$-point PBEsol+$U$~\cite{perdew2008pbesol,dudarev1998} calculations with $U_{\mathrm{eff}}=2.5$~eV on Ti $3d$ were used for two $\sim$7~ps Nos\'e--Hoover NVT AIMD production runs at 900~K and five full-ion conjugate-gradient quenches to $f_{\max}<0.01$~eV\,\AA$^{-1}$. The AIMD time step was 1~fs. Each starting structure was heated from 100 to 900~K over 2~ps and equilibrated at 900~K for 2~ps, followed by approximately 7~ps of production at 900~K. Equilibration and production used a Nos\'e--Hoover NVT thermostat (\texttt{MDALGO=2}), with the Nos\'e mass set by \texttt{SMASS=0} (a 40~fs thermostat period).

\FloatBarrier
\section*{Acknowledgments}
Chengrui Xu conducted this work at the National University of Singapore as a visiting student researcher from Tsinghua University. The authors acknowledge computational resources provided by the National Supercomputing Centre (NSCC) Singapore and the National University of Singapore.
\section*{Data and code availability}
GeoSaddle code will be made available at \url{https://github.com/materialyzeai/geosaddle}. Additional computational details, tables, and figures are provided in the Supplementary Information.
\section*{Author contributions}
C.X.: Methodology, Software, Investigation, Formal analysis, Writing -- original draft.
S.P.O.: Conceptualization, Methodology, Software, Resources, Writing -- review \& editing.

\section*{Declaration of generative AI and AI-assisted technologies}

During this work and the preparation of this manuscript, C.X. used Codex and Gemini to assist with language refinement, literature searches and code development. The authors reviewed and revised the assisted outputs, checked references against their original sources and verified the computational procedures and reported results. The authors made the final decisions on the research design and interpretation and take full responsibility for the accuracy, originality and integrity of the manuscript.

\renewcommand{\bibfont}{\small}
\setlength{\bibsep}{1.5pt}
\bibliographystyle{unsrtnat}
\bibliography{references}
\end{document}


\thispagestyle{plain}
\begin{center}
{\large\bfseries Supplementary Information\par}
\vspace{10pt}
{\LARGE\bfseries GeoSaddle:\\[3pt]
Geometry-Guided Saddle Networks for Periodic Ion Migration\par}
\vspace{14pt}
{\normalsize Chengrui Xu$^1$ and Shyue Ping Ong$^{2,*}$\par}
\vspace{6pt}
{\small $^1$Tanwei College, Tsinghua University\\
$^2$Department of Materials Science and Engineering, National University of Singapore\\[4pt]
$^*$Correspondence: \href{mailto:shyue@nus.edu.sg}{shyue@nus.edu.sg}\par}
\end{center}
\vspace{12pt}
\tableofcontents
\clearpage

\section{Supplementary results}

\subsection{Geometry-guided saddle searches}

All ten events had a near-path saddle in a prior GeoSaddle search. Each seeding method used the same per-event budget (97 dimer attempts in total).

\begin{table}[H]
\centering\small
\caption{\textbf{Saddle searches and potential evaluations on ten events.} A near-path hit is a strict saddle within 0.20~\AA\ of the continuous DFT path. Multiple attempts can hit the same event.}
\label{tab:geometry-guidance}
\begin{tabularx}{\linewidth}{@{}Yrrrrr@{}}
\toprule
& \multicolumn{2}{c}{Attempts} & & \multicolumn{2}{c}{Model calls}\\
\cmidrule(lr){2-3}\cmidrule(l){5-6}
Seed source & Strict & Near-path & Events hit & Preparation & Total\\
\midrule
GeoSaddle & 85 & 72 & 10/10 & 0 & 20,487\\
Random (17) & 71 & 17 & 8/10 & 341 & 44,611\\
Random (29) & 64 & 19 & 7/10 & 341 & 45,535\\
Random (43) & 70 & 18 & 6/10 & 341 & 43,415\\
Ellipsoidal PES grid & 88 & 76 & 10/10 & 143,848 & 161,537\\
\bottomrule
\end{tabularx}
\tabnote{Parentheses give random seeds. Totals cover preparation, dimer search and saddle verification. Preparation includes endpoint relaxation for the random seeds and grid sampling for the ellipsoid.}
\end{table}

\subsection{LiTraj cohorts and auxiliary initialization controls}
\label{app:initialization}

\begin{table}[H]
\centering\small
\caption{\textbf{LiTraj test sets.} The sets share eight hosts and no events. Each LiTraj event specifies the host, endpoints and relative periodic image~\citep{dembitskiy2025litraj}.}
\label{tab:si-cohorts}
\begin{tabularx}{\linewidth}{@{}lrrY@{}}
\toprule
Cohort & Events & Hosts & Selection\\
\midrule
LiTraj-100 & 100 & 93 & Single-Li DFT paths with limited framework motion and a forward barrier below 1~eV.\\
LiTraj-98 & 98 & 92 & Single-ion paths with both directional barriers below 1~eV and $|E_B-E_A|\leq0.30$~eV.\\
\bottomrule
\end{tabularx}
\end{table}

\begin{figure}[H]
\centering
\includegraphics[width=.88\linewidth]{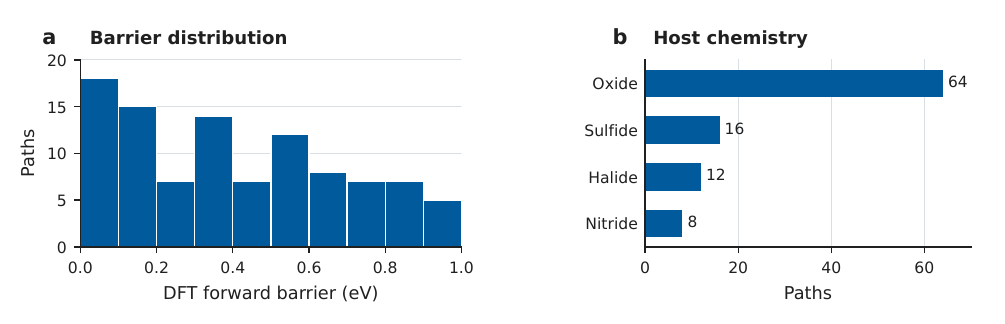}
\caption{\textbf{LiTraj-100 reference paths.} (a) Distribution of forward DFT barriers. (b) Event counts by host chemistry.}
\label{fig:si-initialization}
\end{figure}

Across all 98 LiTraj-98 events, mean final path RMS deviations were 0.145, 0.159 and 0.160~\AA\ for BVSE, CAVD and linear initial paths. With the 21-image UMA force string~\citep{wood2025uma} (\texttt{uma-s-1p2p1}, \texttt{omat}), 60, 45 and 46 paths, respectively, reached 0.05~eV\,\AA$^{-1}$ perpendicular force within 150 iterations.

The H10 control ranked up to five distinct finalists from ten linear, BVSE-interpolated and CAVD paths. The score weighted the highest internal-image energy by 80\% and the mean of the remaining energies by 20\%, relative to the starting endpoint. Up to 20 host-relaxation steps at fixed Li coordinates reduced the mean selected-path RMS from 0.142 to 0.118~\AA.

\subsection{Pathway recovery and transfer to Na}

\begin{figure}[H]
\centering
\includegraphics[width=.88\linewidth]{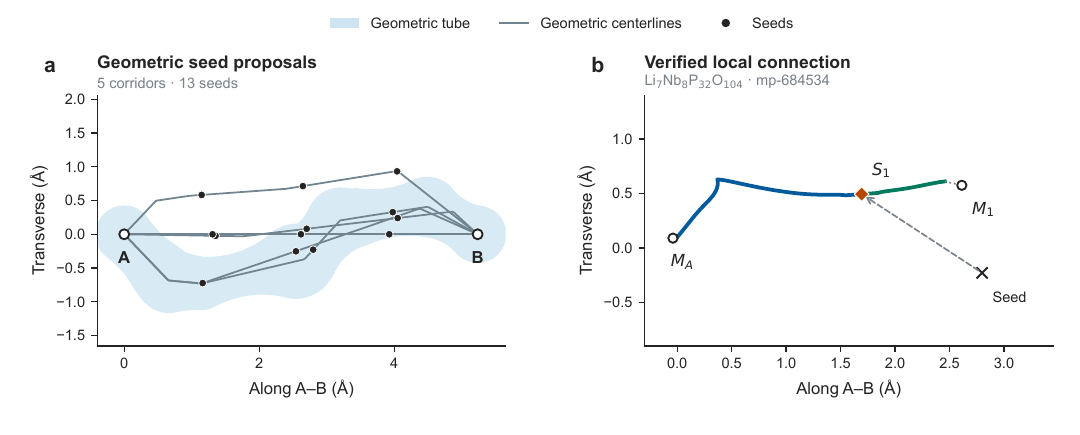}
\caption{\textbf{First saddle connection in \ce{Li7Nb8P32O104} (\texttt{mp-684534}).}
(a) Five channel centerlines, 13 seeds and the projected tube of one route.
(b) The dashed arrow joins the seed to saddle $S_1$. Its two descent branches lead to $M_A$ and $M_1$, the first two minima in main-text Fig.~\ref{main-fig:pathway-neb}a,b.}
\label{fig:si-gpa-example}
\end{figure}

\begin{table}[H]
\centering\small
\caption{\textbf{Eight LiTraj-100 events without a strict $A$--$B$ chain.} All eight contribute to the 100-event recovery denominator. The paired path comparison uses the 92 recovered chains. Figure~\ref{fig:blind100-gallery} shows all 100 events.}
\label{tab:si-blind-outcomes}
\begin{tabularx}{\linewidth}{@{}p{.36\linewidth}Y@{}}
\toprule
Search outcome & Event identifiers\\
\midrule
Endpoints relaxed to one basin &
\texttt{mp-675488\_2\_0\_0\_0\_1}, \texttt{mp-684024\_0\_4\_0\_0\_0}\\
Connection required a borderline saddle &
\texttt{mp-556176\_0\_1\_1\_1\_0}, \texttt{mp-776446\_0\_5\_1\_1\_0}\\
No connection within the search budget &
\texttt{mp-1176627\_2\_7\_0\_1\_0}, \texttt{mp-30249\_0\_5\_0\_0\_0}, \texttt{mp-753520\_2\_6\_0\_0\_0}, \texttt{mp-775160\_2\_2\_0\_0\_1}\\
\bottomrule
\end{tabularx}
\end{table}

\begin{table}[H]
\centering\small
\caption{\textbf{Na saddle-to-path distances in olivine phosphates (\AA).} Each cell contains 24 transition-metal ions. All six hops yielded one strict saddle connecting the endpoints (Fig.~\ref{fig:na-gallery}). Distances are measured to the DFT polyline through five internal images and both endpoints~\citep{petersen2025sodium,petersen2026limitations}.}
\label{tab:na-main}
\begin{tabular}{@{}lrr@{}}
\toprule
Transition metal & 1 Na per cell & 23 Na per cell\\
\midrule
Fe & 0.02086 & 0.04749\\
Mn & 0.03777 & 0.02919\\
Ni & 0.08682 & 0.04726\\
\bottomrule
\end{tabular}
\end{table}

Figure~\ref{fig:na-nasicon} compares the additional Na1--Na2 NASICON hop with its DFT reference~\citep{behara2026nasicons}.

\subsection{Matched NEB refinement on MatPES}

Of 20 events with double-vacancy charge densities, 15 preserved the endpoint basins after MatPES relaxation and yielded converged interior climbing images for all three initial paths.

\begin{table}[H]
\centering\small
\caption{\textbf{Forward barriers for the 15 matched NEB events (eV).} DFT values are from LiTraj. Each MatPES barrier is relative to its relaxed initial endpoint.}
\label{tab:si-barriers}
\begin{tabularx}{\linewidth}{@{}Yrrrr@{}}
\toprule
Event identifier & DFT & BVSE & GeoSaddle & Pathfinder\\
\midrule
\texttt{mp-1020056\_0\_2\_1\_0\_0} & 0.1300 & 0.1415 & 0.1414 & 0.1409 \\
\texttt{mp-1147685\_5\_17\_0\_0\_0} & 0.1319 & 0.1386 & 0.1378 & 0.1385 \\
\texttt{mp-1211142\_7\_0\_0\_0\_1} & 0.6753 & 0.5156 & 0.5157 & 0.5157 \\
\texttt{mp-1211498\_4\_10\_0\_0\_0} & 0.3899 & 0.3542 & 0.3541 & 0.3542 \\
\texttt{mp-1222529\_4\_14\_0\_0\_0} & 0.1620 & 0.1627 & 0.1643 & 0.1623 \\
\texttt{mp-1222582\_1\_10\_0\_0\_0} & 0.2532 & 0.0432 & 0.0422 & 0.0398 \\
\texttt{mp-1276783\_3\_3\_1\_-1\_0} & 0.5505 & 0.5691 & 0.5692 & 0.5691 \\
\texttt{mp-1570562\_5\_7\_0\_0\_0} & 0.5272 & 0.3290 & 0.3290 & 0.3298 \\
\texttt{mp-29077\_0\_3\_1\_0\_0} & 0.2760 & 0.2228 & 0.2222 & 0.2223 \\
\texttt{mp-34038\_4\_4\_0\_1\_0} & 0.1101 & 0.0898 & 0.0902 & 0.0900 \\
\texttt{mp-530262\_10\_12\_0\_0\_0} & 0.2000 & 0.2068 & 0.2071 & 0.2073 \\
\texttt{mp-5418\_7\_5\_1\_0\_0} & 0.6832 & 0.6574 & 0.6578 & 0.6572 \\
\texttt{mp-5853\_0\_1\_0\_0\_0} & 0.6848 & 0.6106 & 0.6109 & 0.6109 \\
\texttt{mp-755571\_0\_1\_0\_1\_0} & 0.7354 & 0.6715 & 0.6719 & 0.6722 \\
\texttt{mp-768967\_12\_22\_0\_0\_0} & 0.2058 & 0.1018 & 0.1026 & 0.1019 \\
\midrule
Mean absolute error relative to DFT & \textemdash & 0.0659 & 0.0660 & 0.0660\\
\bottomrule
\end{tabularx}
\end{table}

\subsection{Periodic defect mechanisms}

\begin{table}[H]
\centering\small
\caption{\textbf{Reference mechanisms used for motif-guided continuation.} The comparison uses site occupancies, intermediate geometries and periodic translations.}
\label{tab:literature-protocols}
\begin{tabularx}{\linewidth}{@{}lYY@{}}
\toprule
System & Published reference & GeoSaddle comparison\\
\midrule
LTP & M1--M2--M1 vacancy hop from PBE NEB~\citep{lu2017ltp} and charge-compensated PBEsol+$U$ CI-NEB~\citep[Fig.~S2]{arjmandi2019latp}. &
One active Li fills the M1 vacancy. Its former site becomes the vacancy for the next search.\\
LATP & Al-associated split-interstitial transfer at $x=0.125$ from PBEsol+$U$ CI-NEB~\citep[Figs.~2--4]{arjmandi2019latp}. PBE CI-NEB resolves the dependence on the local Al environment~\citep[Fig.~4]{pfalzgraf2021latp}. &
Three active Li transfer a split pair between M1 sites through M2 in the fixed Al-decorated host.\\
LPS & PBE AIMD and fixed-cell CI-NEB for one extra Li$^+$ in a charge-compensated $2\times2\times2$ supercell~\citep[Figs.~6--9]{stamminger2020lps}. &
An interstitial placed at $1b$ displaces the lattice Li at $2d$. Four saddle connections resolve the planar exchange through $6k$ and the shallow basin $R$.\\
\bottomrule
\end{tabularx}
\end{table}

\begin{table}[H]
\centering\small
\caption{\textbf{Periodic connections in the finite search networks.} LTP, LATP and LLNOF vectors use the simulation-cell basis. The LPS translation uses the parent-cell basis.}
\label{tab:si-networks}
\begin{tabularx}{\linewidth}{@{}lYY@{}}
\toprule
System & Network or selected chain & Periodic transport\\
\midrule
LTP & 8 states, 22 event records and 5 periodic closures & The two-hop vacancy relay translates by $[0,-1,0]$.\\
LATP & 45 state records and 32 local connections & The four-saddle relay advances by $[0,1,0]$. The largest Li mismatch at a junction is 0.322~\AA.\\
LPS & 5 minima and 4 saddles in one active space & After one parent-cell $b$ translation, the 33 Li sites match with RMS and maximum residuals of 0.051 and 0.165~\AA.\\
LGPS & 132 states and 230 event records, 215 reachable from the initial state & 12 nonzero-winding cycles span two independent translations. The selected relays each carry one net Li along $+c$.\\
LLNOF & 70 states (30 expanded) and 108 verified saddles & 10 nonzero-winding cycles span three independent translations. The selected six-step and three-step relays carry one net Li along $[0,0,1]$ and $[1,-1,0]$.\\
\bottomrule
\end{tabularx}
\end{table}

In LPS the onward barrier from $R$ is $0.18439-0.15689=0.02750$~eV on the common active surface (main-text Fig.~\ref{main-fig:defects}f).

\subsection{LATP configurations sampled by AIMD}

\begin{table}[H]
\centering\small
\caption{\textbf{Five quenched LATP snapshots.} The two RMS columns compare all nine Li with the relaxed start of the corresponding AIMD trajectory and with the closest of 45 GeoSaddle states.}
\label{tab:latp-quenches}
\begin{tabularx}{\linewidth}{@{}llrrrY@{}}
\toprule
Snapshot & Run & Time (ps) & \shortstack{To start\\(\AA)} & \shortstack{To network\\(\AA)} & Network assignment\\
\midrule
Q00 & S0 & 2.920 & 0.021 & 0.131 & Starting basin\\
Q01 & S0 & 4.838 & 0.788 & 0.112 & Off-chain state\\
Q02 & S0 & 6.665 & 1.186 & 0.230 & Off-chain state\\
Q03 & S1 & 1.320 & 0.822 & 0.102 & Off-chain state\\
Q04 & S1 & 2.340 & 0.573 & 0.525 & Unmatched\\
\bottomrule
\end{tabularx}
\tabnote{The host RMS residuals after alignment were 0.069--0.074~\AA. In Q04 the largest individual Li mismatch was 1.542~\AA.}
\end{table}

\subsection{LGPS relay energetics and site assignments}

\begin{figure}[H]
\centering
\includegraphics[width=.84\linewidth]{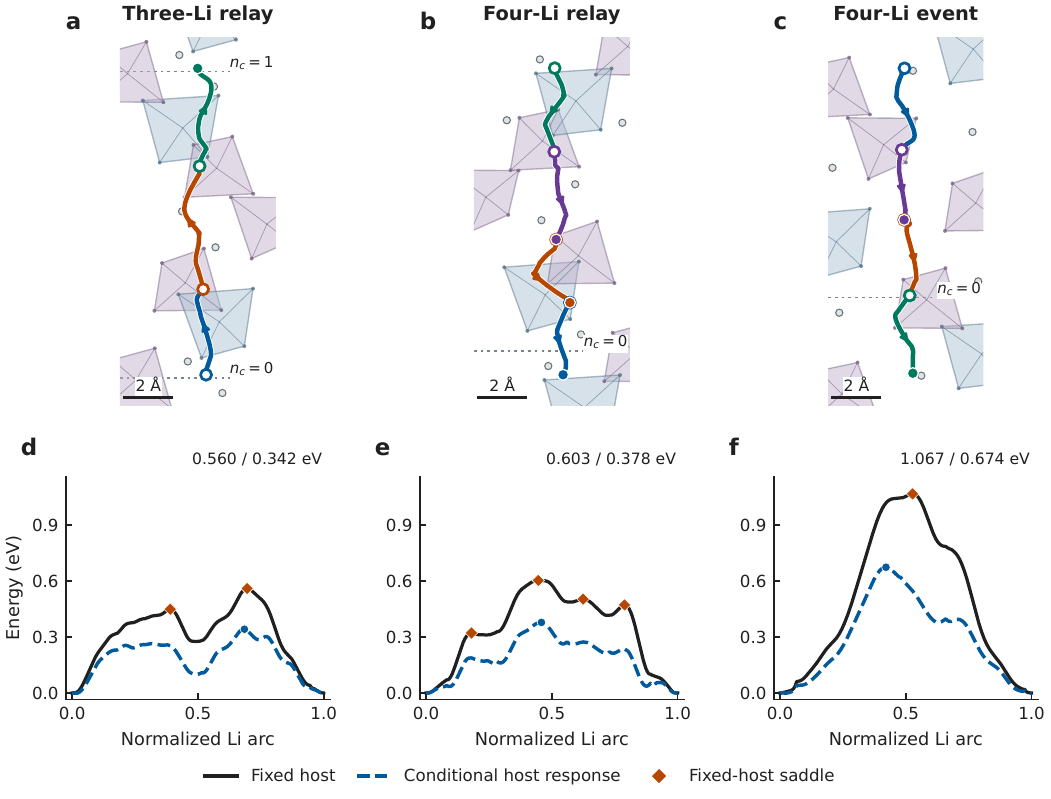}
\caption{\textbf{Three collective pathways in LGPS.} (a--c) The three-Li relay, four-step four-Li relay and single-saddle four-Li event. Colors track individual Li and dotted lines mark $c$-cell boundaries. (d--f) MatPES profiles with the host fixed (solid) and relaxed at fixed Li coordinates (dashed). Diamonds mark fixed-host saddles.}
\label{fig:si-lgps-matpes}
\end{figure}

\begin{table}[H]
\centering\small
\caption{\textbf{LGPS profile peaks above state $A$ (eV).} Each row uses the energy of $A$ under the same potential and relaxation constraints as the profile.}
\label{tab:lgps-main}
\begin{tabularx}{\linewidth}{@{}Yrrr@{}}
\toprule
& \shortstack{Three-Li relay\\(two saddles)} & \shortstack{Four-Li relay\\(four saddles)} & \shortstack{Four-Li event\\(one saddle)}\\
\midrule
Fixed-host MatPES & 0.560 & 0.603 & 1.067\\
MatPES, host relaxed at fixed Li & 0.342 & 0.378 & 0.674\\
Full-ion DFT CI-NEB & 0.346 & 0.116 & \textemdash\\
\bottomrule
\end{tabularx}
\tabnote{DFT refinement covered the two relays. The conditional MatPES peaks are evaluated along fixed Li trajectories.}
\end{table}

\begin{table}[H]
\centering\small
\caption{\textbf{DFT CI-NEB segments of the two LGPS relays.} Peak energies are relative to $A$, local barriers to each segment's initial minimum. RMS values average the internal-image distances to the GeoSaddle polyline. Relay weight is the designated ions' share of the squared Li displacements.}
\label{tab:si-lgps-six}
\begin{tabular}{@{}llrrrrrr@{}}
\toprule
Relay & Step & Peak & Local & Max. force & Relay RMS & All-Li RMS & Relay weight\\
 & & (eV) & (eV) & (eV/\AA) & (\AA) & (\AA) & (\%)\\
\midrule
Three-Li & 1 & 0.346 & 0.346 & 0.04960 & 0.151 & 0.131 & 88.5 \\
Three-Li & 2 & 0.343 & 0.240 & 0.04999 & 0.139 & 0.103 & 90.3 \\
Four-step & 1 & 0.086 & 0.086 & 0.04692 & 0.144 & 0.154 & 87.9 \\
Four-step & 2 & 0.083 & 0.011 & 0.04180 & 0.300 & 0.271 & 92.9 \\
Four-step & 3 & 0.108 & 0.059 & 0.04964 & 0.344 & 0.328 & 73.0 \\
Four-step & 4 & 0.116 & 0.036 & 0.04843 & 0.141 & 0.225 & 70.4 \\
\bottomrule
\end{tabular}
\end{table}

The endpoint exchanges are $5\to14\to16\to5[001]$ and $4\to10\to19\to15\to4[001]$ (zero-based Li indices). The experimental site registry~\citep{kuhn2013lgps} was aligned using the 30 framework atoms (0.269~\AA\ RMS). Both relays follow Li1/Li3 channels.

\begin{figure}[H]
\centering
\includegraphics[width=.89\linewidth]{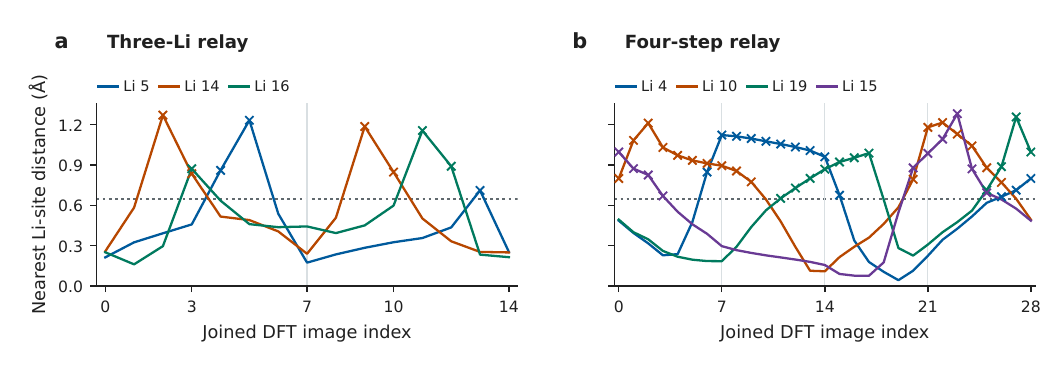}
\caption{\textbf{Distance to the nearest experimental Li site in LGPS.} Each curve follows one relay ion through the joined DFT bands for (a) the three-Li and (b) the four-step relay. Crosses mark distances above 0.65~\AA.}
\label{fig:si-lgps-registry}
\end{figure}

\section{Computational details}
\label{app:experimental-details}

\subsection{Geometry construction and seed placement}
\label{si:geometry}

The lifted Cartesian grid used $h=0.12$~\AA, a focal-distance stretch of 2.0, a 0.50~\AA\ halo and 26-neighbor connectivity. Each edge had lower clearance bound $w^-_{uv}=\min(c_u,c_v)-d_{uv}/2$ for edge length $d_{uv}$. The upper estimate used node capacities and the voxel cover radius $\sqrt{3}h/2$. Continuous connectors attached both endpoints.

We used tabulated ionic radii $\rho_i$ where available and stored BVSE radii otherwise. Opposite-charge neighbors within 4.0~\AA\ contribute to $K_i$ at a bond-valence cutoff of 0.05, giving
\[
s_{ij}=\exp[(R_{ij}-d_{ij})/b],\qquad
K_i=\sum_j s_{ij}/d_{ij}^{\,2},\qquad
\rho_i^{\mathrm{eff}}=\max\!\left(0,\rho_i-\frac{u_{\max}}{1+K_i/K_{\mathrm{ref}}}\right).
\]
Here $R_{ij}$ is the bond-valence reference distance, $b=0.31$~\AA, $u_{\max}=0.25$~\AA, and $K_{\rm ref}$ is the 90th percentile of positive $K_i$. The host coordinates remain fixed.

\begin{algorithm}[H]
\small
\caption{Geometry-guided seed construction}
\label{alg:geometry}
\begin{algorithmic}[1]
\Require Host obstacles and lifted endpoints $A,B$
\State Construct the clearance graph and compute widest-path capacities from both endpoints on its maximum spanning forest.
\State At each of 12 aperture thresholds, find a shortest admissible $A$--$B$ path. Block previous bottlenecks to seek up to three alternatives.
\State Sample shortcut chords at 0.01~\AA\ and resample centerlines at 0.04~\AA. Limit path length to $1.8|B-A|$ and merge duplicate curves at 0.25~\AA\ RMS over 64 arc-length samples.
\State On routes with minimum clearance $\geq0.40$~\AA, place tangent seeds at $1/4$, $1/2$ and $3/4$ of the arc length. Include the direct $A$--$B$ midpoint with its chord direction.
\State \Return the routes and the seed set after merging positions within 0.20~\AA.
\end{algorithmic}
\end{algorithm}

\subsection{Stationary-point verification}
\label{si:verification}

The active-space Hessian used symmetrized central force differences at $\pm0.02$~\AA. The dimer separation was $10^{-4}$~\AA. Endpoint searches allowed four rotations per translation and a maximum step of 0.03~\AA.

\begin{table}[H]
\centering\small
\caption{\textbf{Active spaces and dimer settings.} The Hessian index counts eigenvalues below the negative of the listed tolerance. $n$ is the number of atoms in the local Li--host active space.}
\label{tab:si-active-spaces}
\begin{tabularx}{\linewidth}{@{}Yrrrr@{}}
\toprule
Search & Coordinates & Max. steps & \shortstack{Force target\\(eV/\AA)} & \shortstack{Index tolerance\\(eV/\AA$^2$)}\\
\midrule
Single-ion endpoints & 3 & 180 & 0.02 & 0.02\\
Local Li--host refinement & $3n$ & 220 & 0.03 & 0.02\\
LATP three-Li search & 9 & 240 & 0.02 & 0.02\\
LPS common-space validation & 165 & 600 & 0.001 & 0.002\\
LGPS all-Li search & 60 & 600 & 0.001 & 0.002\\
LLNOF all-Li search & 27 & 600 & 0.001 & 0.002\\
\bottomrule
\end{tabularx}
\end{table}

Strict single-ion saddles required dimer convergence~\citep{henkelman1999dimer}, active force $\leq0.03$~eV\,\AA$^{-1}$, index one at the tolerance in Table~\ref{tab:si-active-spaces}, and energy decreases at both $\pm0.10$~\AA\ mode probes. FIRE descents~\citep{bitzek2006fire} started at $\pm0.15$~\AA. Both minima had to reach 0.02~eV\,\AA$^{-1}$ within 400 steps and pass the zero-index check.

LPS, LGPS and LLNOF minima used a 0.0005~eV\,\AA$^{-1}$ force target. Three LATP relay saddles passed both mode probes. The fourth rose by 2.08~meV on one probe and descended to verified minima on both sides.

\subsection{Matching, recovery and periodic exploration}
\label{si:network}

Endpoint searches clustered minima within 0.30~\AA\ and matched them to $A$ or $B$ within 0.75~\AA. Chains were ranked by maximum saddle energy, then by edge count and saddle identifier. Recovery added radical-Voronoi and string proposals, linked intermediate minima to depth two and split index-two peaks along transverse modes. LiTraj-100 also used a near-critical short-hop screen.

\begin{figure}[H]
\centering
\includegraphics[width=.94\linewidth]{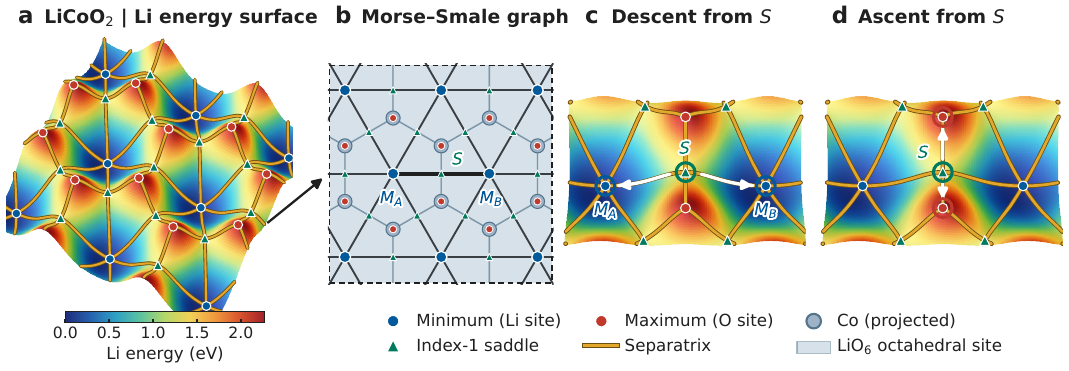}
\caption{\textbf{Stationary points in the Li-layer plane of O3-\ce{LiCoO2}.} One Li moves on the MatPES surface with the \ce{CoO2} host fixed and other Li removed. (a,b) Energy surface and in-plane network. (c,d) Descending paths connect saddle $S$ to minima $M_A$ and $M_B$. Ascending paths separate the basins.}
\label{fig:morse-smale}
\end{figure}

Periodic matching minimized the sum of Li distances in LATP and squared distances within each species in LPS, LGPS and LLNOF. The latter minima required RMS $\leq0.03$~\AA, maximum residual $\leq0.10$~\AA\ and $|\Delta E|\leq0.001$~eV. LGPS and LLNOF saddle deduplication used 0.015~\AA\ RMS and 0.05~\AA\ maximum residual.

\begin{algorithm}[H]
\small
\caption{Periodic exploration of a mobile-ion sublattice}
\label{alg:exploration}
\begin{algorithmic}[1]
\Require The relaxed starting minimum, active Li coordinates and the potential
\State Initialize the graph and a first-in, first-out queue with the starting minimum.
\While{the queue and budget allow further searches and the target winding has not been found}
\State Generate proposals from the next state's occupancies. Verify the saddles and register both products with their atom permutations and lattice-image shifts.
\State Queue newly identified minima. Target disconnected basins with ridge proposals and rebuild the local channel skeleton if expansion stalls.
\State Compose the edge permutations and image shifts around cycles reachable from the start. Test for nonzero winding along the target direction.
\EndWhile
\State \Return the verified graph and periodic cycles.
\end{algorithmic}
\end{algorithm}

LATP used the lowest-energy of 16 Al substitutions and a 0.45~\AA\ maximum Li residual for periodic closure. The 97-atom LPS supercell contained one extra $1b$ Li at $(0,0,1/4)$ in the $2\times2\times2$ cell~\citep{stamminger2020lps}. Translation by $[0,1,0]$ in the parent cell corresponds to $[0,\tfrac12,0]$ in the supercell.

The LGPS $c$-axis search used a 0.5~\AA\ skeleton grid, with limits of 48 expanded states and 160 screened proposals per state. The root received 48 dimer attempts and later states up to 16, plus up to two ridge proposals per state. The three-Li relay closed during queue expansion and the four-step relay during targeted continuation.

For LLNOF ($Fd\bar{3}m$, $Z=8$, $a=10.44429$~\AA\ at $x=0.875$)~\citep{aimi2024llnof}, single-point MatPES screening of all 1,330 symmetry-distinct $16d$ orderings and full-ion relaxation of the lowest 12 candidates selected $S_{000}$ (\ce{Li9La5Nb16O48F8}, 6.65~meV below the next ordering). Three-dimensional exploration used the same 0.5~\AA\ skeleton grid, proposal limits and dimer budgets as LGPS until reaching transport rank $r=3$ at 60 states and 87 saddles. Continuation from the eight unexpanded states within 0.90~eV of $S_{000}$ completed the 70-state, 108-saddle graph.

\subsection{Host relaxation and energetic refinement}
\label{si:refinement}

Local-host LiTraj refinement included non-Li neighbors within 4.0~\AA\ of the saddle. The refined saddle had to retain index one and remain within 0.75~\AA\ of the fixed-host Li position. Conditional LTP, LATP, LGPS and LLNOF profiles relaxed the host at fixed Li coordinates relative to the identically relaxed initial state. For LGPS and LLNOF the force target was 0.001~eV\,\AA$^{-1}$ and the maximum step was 0.03~\AA.

\paragraph{Matched NEB calculations.}
Pathfinder~\citep{rong2016pathfinder} used the double-vacancy charge density ($h=3.0$, $k=0.17$). After full-ion endpoint relaxation to 0.01~eV\,\AA$^{-1}$, the basin-retention limits were 1.0~\AA\ for the mobile ion and 0.50~\AA\ for any host atom. The three 21-image MatPES bands shared the relaxed endpoints, linearly interpolated hosts and a spring constant of 0.1~eV\,\AA$^{-2}$. Ordinary and climbing stages used force targets of 0.05 and 0.02~eV\,\AA$^{-1}$, step limits of 300 and 600, and a maximum step of 0.10~\AA.

\paragraph{Full-ion MatPES refinement.}
For the six-step $+c$ relay in LLNOF, all 86 atoms were relaxed at the six junctions to 0.003~eV\,\AA$^{-1}$ at fixed cell and connected by six 11-image MatPES CI-NEB bands ($k=0.1$~eV\,\AA$^{-2}$, maximum step 0.08~\AA). Ordinary and climbing stages used force targets of 0.05 and 0.03~eV\,\AA$^{-1}$ with step limits of 350 and 450, and per-ion displacement weights followed the same squared-step definition as in LGPS.

\paragraph{Full-ion DFT refinement.}
Full-ion PBE PAW bands in VASP~\citep{perdew1996pbe,blochl1994paw,kresse1996vasp} used six internal images, a 520~eV cutoff, fixed cells, \texttt{SPRING=-5}, \texttt{IBRION=1} and \texttt{POTIM=0.2}. For \ce{Li5BO4}, all three initial paths shared the relaxed endpoints and a $\Gamma$-centered $3\times3\times3$ mesh ($10^{-7}$~eV electronic tolerance, 0.03~eV\,\AA$^{-1}$ force target). LGPS used a $\Gamma$-centered $2\times2\times2$ mesh and \texttt{Li\_sv}, \texttt{Ge\_d}, \texttt{P} and \texttt{S} potentials. Junctions were relaxed to 0.01~eV\,\AA$^{-1}$ and shared between adjacent bands. All six bands reached 0.05~eV\,\AA$^{-1}$.

For Table~\ref{tab:si-lgps-six}, each internal DFT image was aligned by host translation and projected onto the continuous GeoSaddle path using joint RMS displacement at fixed Li identity. The relay weight sums the following per-ion contributions over the relay ions, where
\[
 w_i=\frac{\sum_k\|\mathbf r_i^{k+1}-\mathbf r_i^k\|^2}
           {\sum_{j\in\mathrm{Li}}\sum_k\|\mathbf r_j^{k+1}-\mathbf r_j^k\|^2}.
\]

\subsection{Cohorts and geometric error metrics}
\label{si:evaluation}

The ten-case subset required endpoint separation $\geq4$~\AA, path elongation $\geq1.12$ and transverse deviation $\geq0.80$~\AA. Random seeds used displacements of 0.3, 0.6 or 0.9~\AA\ about relaxed endpoints and their directions as initial modes. Grid seeds used path tangents in an ellipsoid with foci $A,B$, eccentricity 0.70 and spacing $\leq0.20$~\AA.

Path RMS used 21 equal arc-length samples on the predicted path and the nearest points on continuous DFT segments after endpoint-compatible symmetry alignment. Coverage was the fraction of 101 DFT arc-length samples within 0.20~\AA\ of the recovered chain.

For each strict chain, we subtracted the squared BVSE distance from the squared GeoSaddle distance at each of the 21 fractions. The nominal 95\% interval used the 2.5th and 97.5th percentiles of 20,000 paired circular block-bootstrap means (block length five, 21 samples per replicate)~\citep{politis1992circular}.

\subsection{LATP first-principles and trajectory comparisons}
\label{si:latp-aimd}

The 145-atom LATP cell used spin-polarized PBEsol+$U$ ($U_{\rm eff}=2.5$~eV on Ti $3d$), $\Gamma$-point sampling, a 520~eV cutoff and 0.05~eV Gaussian smearing~\citep{perdew2008pbesol,dudarev1998}. S0 was a DFT-relaxed GeoSaddle state. S1 was relaxed after one three-Li displacement from S0. Each trajectory used 2~ps of heating (100--900~K), 2~ps of equilibration and about 7~ps of production at 900~K, with a 1~fs step. Electronic tolerances were $10^{-5}$~eV for AIMD and $10^{-6}$~eV for quenching. All atoms relaxed to below 0.01~eV\,\AA$^{-1}$ at fixed cell.

After host-translation alignment, each quench was matched over all nine Li and the three symmetries of the Al-decorated framework. The 2,499 frames per trajectory were weighted by their 1 or 10~fs output intervals. Occupancies on a periodic $48^3$ grid were normalized per trajectory and smoothed by a one-voxel Gaussian for display. A single framework symmetry was used for route distances throughout the S1 event (0.210--2.510~ps).

\clearpage
\section{Migration path atlas}
\label{app:paths}

\subsection{LiTraj-100}
\label{app:paths-blind}

\pdfximage{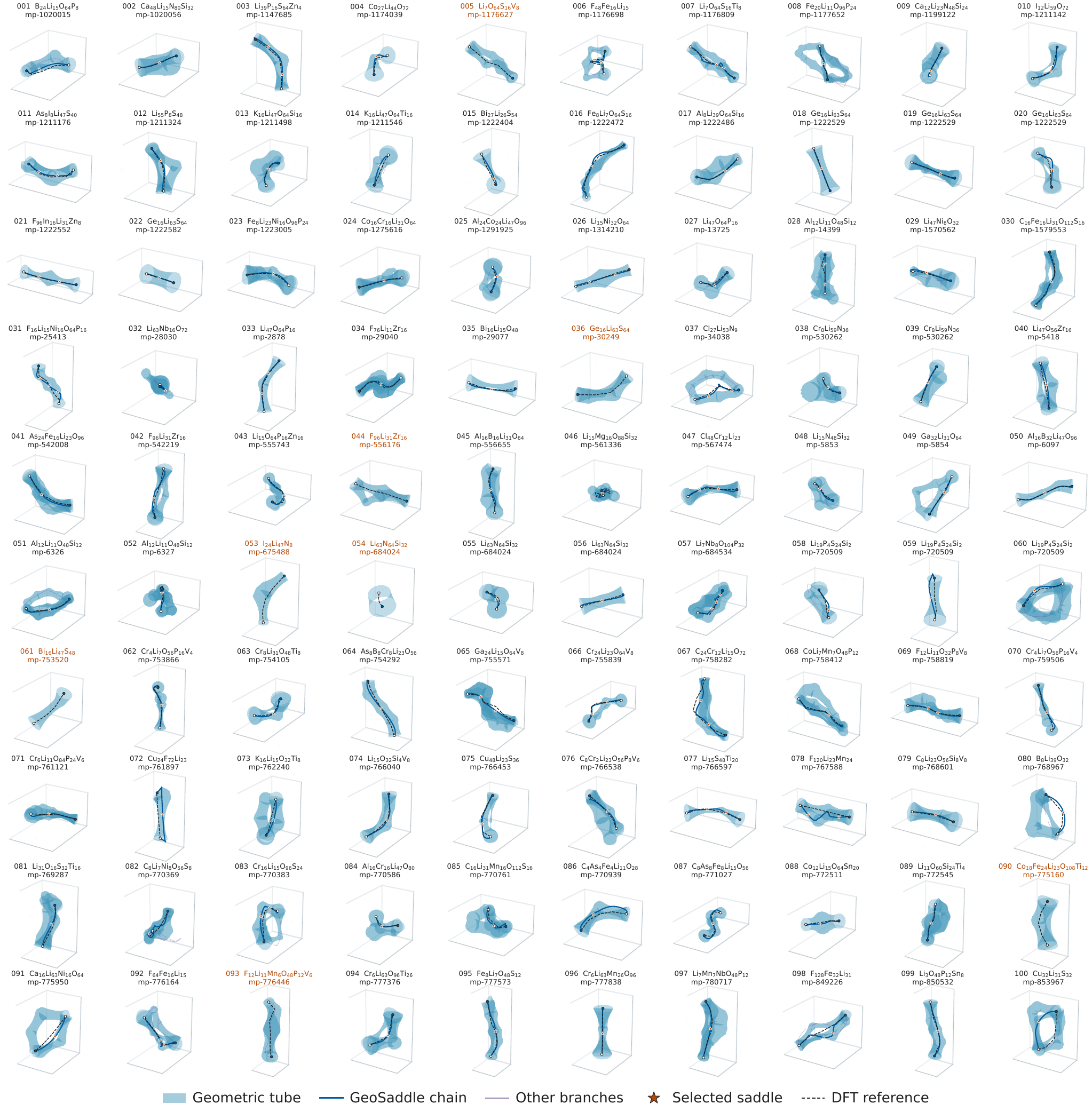}
\edef\atlaslastpage{\the\pdflastximagepages}
\edef\atlasfirstplate{\ifnum\atlaslastpage>1 2\else 1\fi}
\begin{figure}[H]
\centering
\includegraphics[page=\atlasfirstplate,width=\linewidth]{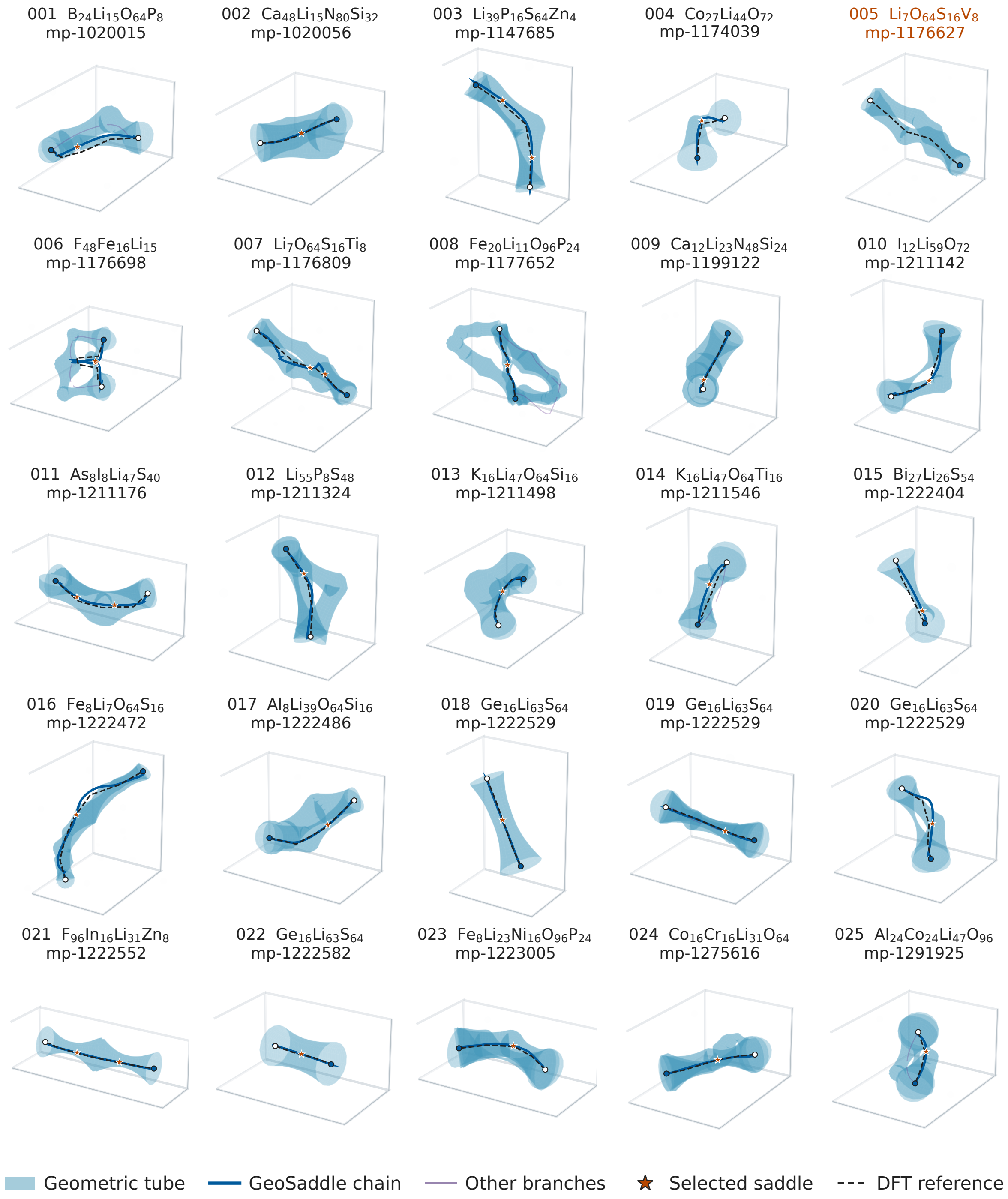}
\caption{\textbf{LiTraj-100 migration paths.} Light blue marks geometric tubes, dark blue the selected chains and dashed black the DFT paths. Stars mark selected saddles. Orange titles identify the eight events without a strict chain.}
\label{fig:blind100-gallery}
\end{figure}

\newcommand{\atlascontinuedplate}[1]{%
  \clearpage
  \begin{figure}[H]
  \ContinuedFloat
  \centering
  \includegraphics[page=#1,width=\linewidth]{gallery/blind100_overview.pdf}
  \caption[]{\textbf{LiTraj-100 migration paths (continued).}}
  \end{figure}%
}
\ExplSyntaxOn
\int_step_function:nnN { \atlasfirstplate + 1 } { \atlaslastpage } \atlascontinuedplate
\ExplSyntaxOff

\subsection{Na transfer}
\label{app:paths-na}

\begin{figure}[H]
\centering
\includegraphics[width=.94\linewidth]{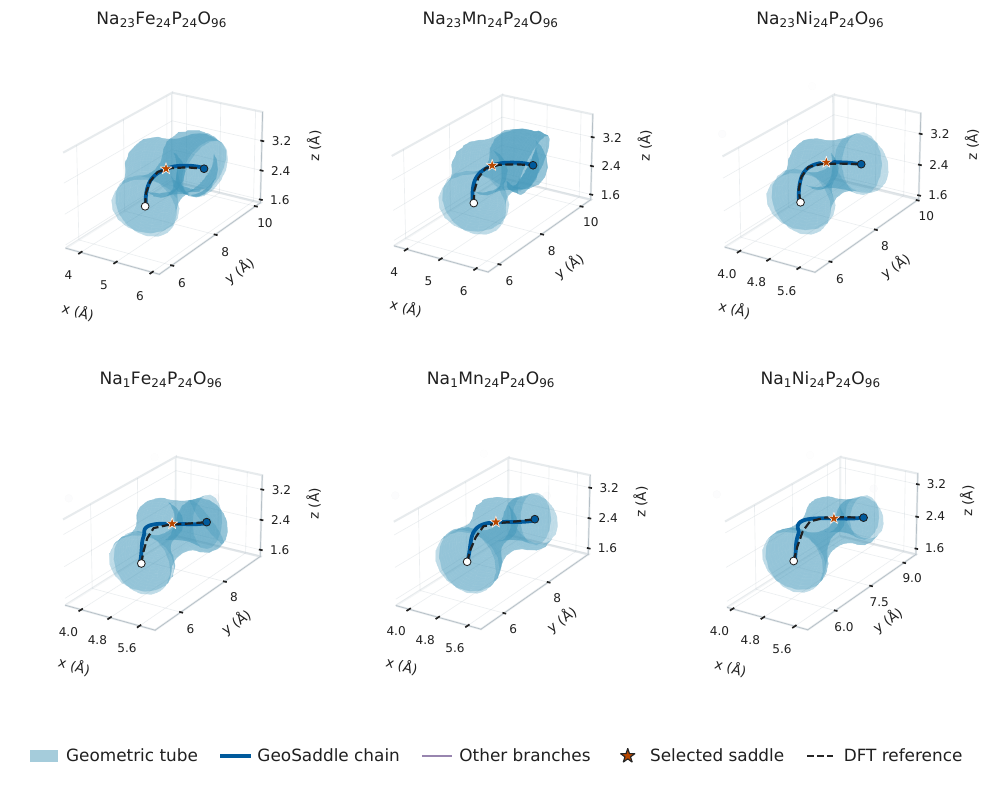}
\caption{\textbf{Six Na hops in olivine phosphates.} Columns show Fe, Mn and Ni. The upper and lower rows contain 23 and one Na per 24 transition-metal ions, respectively. Curves and symbols follow Fig.~\ref{fig:blind100-gallery}. Saddle-to-path distances are given in Table~\ref{tab:na-main}.}
\label{fig:na-gallery}
\end{figure}

\begin{figure}[H]
\centering
\includegraphics[width=.72\linewidth]{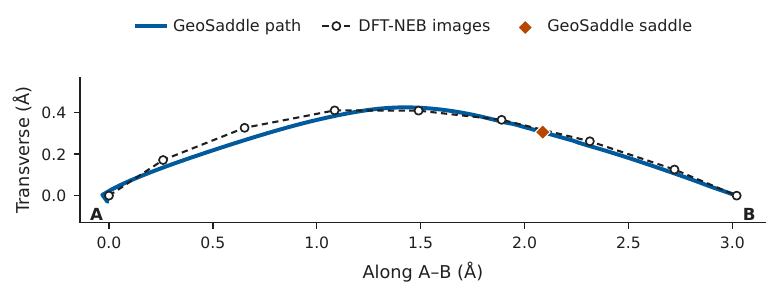}
\caption{\textbf{Na1--Na2 migration in NASICON \ce{Na4MnCr(PO4)3}.} Open circles are the nine DFT-NEB images~\citep{behara2026nasicons}. The blue curve is the GeoSaddle path and the orange diamond its saddle in the same projection. The three-dimensional path RMS and saddle-to-DFT-path distance are 0.027 and 0.012~\AA, respectively.}
\label{fig:na-nasicon}
\end{figure}

\clearpage
\bibliographystyle{unsrtnat}
\begingroup
\footnotesize
\setlength{\bibsep}{1.5pt}
\setlength{\parskip}{0pt}
\renewcommand{\bibsection}{}
\begin{multicols}{2}[\section*{\refname}]
\bibliography{references}
\end{multicols}
\endgroup